\documentclass[a4paper,11pt]{article}
\usepackage{amssymb,graphicx,amsmath,color}

\newcommand{\sect}[1]{\setcounter{equation}{0}\section{#1}}

\input{tcilatex}

\newcommand{\bc}{\begin{center}}
\newcommand{\ec}{\end{center}}
\def\ba#1{\begin{array}{#1}\displaystyle}
\newcommand{\ea}{\end{array}}

\newcommand{\beq}{\begin{equation}}
\newcommand{\eeq}{\end{equation}}
\newcommand{\beqa}{\begin{eqnarray}}
\newcommand{\eeqa}{\end{eqnarray}}
\newcommand{\no}{\nonumber}
\newcommand{\n}{\nonumber\\}
\newcommand{\bi}{\begin{itemize}}
\newcommand{\ei}{\end{itemize}}

\def\sect#1{\section{#1}\setcounter{equation}{0}}

\def\lt#1{\left#1}
\def\rt#1{\right#1}
\def\t#1{\tilde{#1}}
\def\h#1{\hat{#1}}
\def\b#1{\bar{#1}}
\def\frc#1#2{\frac{#1}{#2}}

\newcommand{\bra}{\langle}
\newcommand{\ket}{\rangle}

\newcommand{\R}{{\mathbb{R}}}
\newcommand{\C}{{\mathbb{C}}}
\newcommand{\CP}{{\mathbb{C}{\rm P}}}

\newcommand{\Or}{{\cal O}}

\newcommand{\ep}{\epsilon}

\newcommand{\Tr}{{\rm Tr}}

\newcommand{\End}{{\rm End}}

\newcommand{\tS}{{\tt P}}

\def\cvec#1{\mathbf{#1}}

\begin{document}

\setcounter{page}{0} \topmargin0pt \oddsidemargin0mm \renewcommand{%
\thefootnote}{\fnsymbol{footnote}} \newpage \setcounter{page}{0}
\begin{titlepage}
\vspace{0.2cm}
\begin{center}
{\Large {\bf Entanglement in permutation symmetric states, fractal dimensions, and geometric quantum mechanics}}


\vspace{0.8cm} {\large \text{Olalla A.~Castro-Alvaredo$^{\bullet}$
and Benjamin Doyon$^{\circ}$}}

\vspace{0.2cm}
{$^{\bullet}$  Department of Mathematical Science, City University London, \\
Northampton Square, London EC1V 0HB, UK}\\
{$^{\circ}$  Department of Mathematics, King's College London,
\\
Strand, London WC2R 2LS, UK}
\end{center}
\vspace{1cm} We study the von Neumann and R\'enyi bipartite entanglement entropies in the thermodynamic limit of many-body quantum states with spin-$s$ sites, that possess full symmetry under exchange of sites.
It turns out that there is essentially a one-to-one correspondence between such thermodynamic states and probability measures on $\CP^{2s}$.
  Let a measure be supported on a set of possibly fractal real dimension $d$ with respect to the Study-Fubini metric
  of $\CP^{2s}$. Let $m$ be the number of sites in a subsystem of the bipartition. We give evidence that in the limit $m\to\infty$, the entanglement
   entropy diverges like $\frc d2\log m$. Further, if the measure is supported on a submanifold of $\CP^{2s}$ and can be
   described by a density $f$ with respect to the metric induced by the Study-Fubini metric, we give evidence that the correction
   term is simply related to the entropy associated to $f$: the geometric entropy of geometric quantum mechanics.
    This extends results obtained by the authors in a recent letter where the spin-$\frac{1}{2}$ case was considered.
     Here we provide more examples as well as detailed accounts of
      the ideas and computations leading to these general results. For special choices of the state in the spin-$s$ situation, we recover the scaling behaviour
      previously observed by Popkov et al., showing that their result is but a special case of a more general scaling law.
 \vfill{
\hspace*{-9mm}
\begin{tabular}{l}
\rule{6 cm}{0.05 mm}\\
$^\bullet \text{o.castro-alvaredo@city.ac.uk}$\\
$^\circ \text{benjamin.doyon@kcl.ac.uk}$\\
\end{tabular}}

\renewcommand{\thefootnote}{\arabic{footnote}}
\setcounter{footnote}{0}

\end{titlepage}
\newpage
\sect{Introduction}
The study of the entanglement entropy of extended quantum systems such as quantum spin chains has attracted much attention in recent years (see e.g. the
recent special issue \cite{specialissue}). The entanglement entropy is an interesting quantity for many reasons. Among other features, it gives
information about the quantum state of a system, particularly about the amount of entanglement that it may store, and
it exhibits ``universal" behaviour, for example in systems at or near
conformal critical points.

There are different types of entanglement entropy but a widely studied quantity is the von Neumann entropy. Formally, it may
be defined as follows:
 consider a quantum system which we partition into two susets $A$ and $\bar{A}$ and define its
Hilbert space ${\cal H} = {\cal
H}_A\otimes{\cal H}_{\b{A}}$. Suppose the system is in a pure state $|\psi\rangle$. The
bipartite entanglement entropy $S$ is the von Neumann entropy
associated to the reduced density matrix $\rho_A$ of the subsystem
$A$,
\begin{equation}
    \rho_A = \text{Tr}_{{\cal H}_{\b{A}}}(|\psi\rangle\langle \psi|)\,,
\end{equation}
given by
\begin{equation}\label{vneu}
   S=-\text{Tr}_{\mathcal{H}_A} (\rho_A \log \rho_A).
\end{equation}
A generalization of the von Neumann entropy is the R\'enyi entropy which is defined as
\begin{equation}
    S_n=\frac{\log (\Tr_{\mathcal{H}_A}(\rho_A^n))}{1-n},
    \label{renyi}
\end{equation}
and depends on an additional parameter $n$. In the limit $n\rightarrow 1$ it equals the von Neumann entropy. The quantity $\rho_A^n$ may be interpreted as
the reduced density matrix of subsystem $A$ in a `replica' theory consisting of $n$ copies of the original model. Hence, one often refers to
$n$ as a `number of copies'.

For quantum systems at conformal criticality, it is well know that the von-Neumann  and R\'enyi entropies
of a large block of spins of size $m$ diverge as
\begin{equation}
\label{known}
    S\sim \frac{c}{3}\log m \qquad \text{and} \qquad S_n \sim \frac{c}{6}\left(1+n^{-1}\right)\log m,
\end{equation}
where $c$ is the central charge of the conformal field theory which describes the quantum critical point \cite{HolzheyLW94,Calabrese:2004eu,Calabrese:2005in}.
The first equation in (\ref{known}) is arguably the best known result in this field of research and one that has been successfully tested
 for a great variety of theories, both through numerical and analytical methods.

Less known is the fact that the entanglement entropy can also reveal very interesting information about the state of quantum systems away
from conformal criticality.
An important body of work exists where the entropy of quantum spin chains with random interactions
has been studied (see e.g. the review \cite{random}) leading to scaling laws for the entropy similar to (\ref{known}) but where the coefficient
$\frac{c}{3}$ is often replaced by an irrational number. Also, a number of relatively recent works \cite{pop1,pop2,vidal,ravanini,permutation,fractal}
involving for instance non-critical quantum spin chains (i.e.~spin chains whose thermodynamic limit is not described by conformal field theory) or models with long-range interactions, have found a variety of
behaviours for the entropy of large blocks of spins. Our own work \cite{fractal} has, for the first time, provided a geometric interpretation as well as
a generalization of some of those results for the spin-$\frac{1}{2}$ case. In \cite{fractal} we studied the entanglement entropy of an infinite
spin-$\frac{1}{2}$
chain whose ground state is an infinite linear combination of basic permutation symmetric, zero entropy  states.
We found that
both the von Neumann and R\'enyi entropies of a large block of spins diverge logarithmically with the size of the block $m$ as
\begin{equation}\label{ent}
    S, S_n\sim \frac{d}{2}\log m  + O(1)\quad \text{with} \quad d\in[0,2].
\end{equation}
In particular, and in contrast to the results in one-dimensional conformal critical systems, we obtain the same behaviour for the von Neumann and R\'enyi entropies, as well as the single-copy entropy \cite{SC1,SC2}, that is, the limit $\lim_{n\to\infty}S_n$ (we will see that the $O(1)$ term is finite in this limit).
 
A particular choice of the state $|\psi\rangle$ (known as a Dicke state \cite{dicke}) leading to $d=1$ is the ground state of the ferromagnetic XXX chain, and was studied in \cite{pop1,permutation}. Also, in \cite{vidal}, this and other permutation symmetric states (certain linear combinations of Dicke states) were studied as ground states of the Lipkin-Meshkov-Glick (LMG) model \cite{LMG,MGL,GLM}\footnote{Note that it was found in \cite{vidal2} that in the LMG model the same critical behaviour is observed for the von Neumann and single-copy entropies.}.
In \cite{fractal} we argued that the quantity $d$ can be any real number in the given interval and that it represents a dimension
which characterizes the geometry (fractal or otherwise) of the quantum state. For spin-$\frac{1}{2}$ all allowed quantum states of
the system have support on the Bloch sphere of geometric dimension $d=2$. In other words, all states of the system can be expressed as
linear combinations of states $|\psi_{\mathbf{v}}\rangle$ labeled by a vector $\mathbf{v}$ on the Bloch sphere and as the length of the chain tends to infinite,
every vector in the Bloch sphere corresponds to a basis state of the system. In particular, we may consider a quantum state associated to the `embedding' of a fractal object in the Bloch sphere,
such as the Cantor set shown
in Fig.~1 of Section 6.1 and recover the fractal dimension of such object by computing the bipartite entanglement entropy.

In the current work we put the results above on firmer mathematical footing as well as showing
that the method employed in \cite{fractal} can been successfully generalized to higher spins $s$ and general permutation
symmetric states. We identify the geometry of the support of the states of such a system and find that the Bloch sphere is naturally
generalised to the projective complex space $\CP^{2s}$.
We show that for such states the entanglement entropy behaves as in (\ref{ent}) with $d \in [0,4s]$, and that the explicit form of the correction term $O(1)$ in (\ref{ent}), which is given in the next section, naturally singles out the Study-Fubini metric of $\CP^{2s}$. The present work provides mathematically precise statements (see below Theorems \ref{th1} and \ref{th2}, and Conjecture \ref{th3}), as well precise and compelling derivations (if not rigorous proofs) of them.

The paper is organized as follows: Due to the technical nature of some of our results we have decided to start the paper by introducing the notation and main terminology we will be using in the remainder of the work. We then summarise, discus and set our main findings in context, comparing them when possible to previous existing work. We do all of this in Section \ref{sectmain}. In Section \ref{sectth2} we provide a proof of one of our main results (Theorem 2 stated in Section \ref{sectmain}). This result is a general formula for the R\'enyi entropy based on the use of cyclic permutation operators introduced in previous work by the authors. In Section \ref{sectThermo} we consider the Thermodynamic limit of permutation symmetric states, that is we find a suitable representation for permutation symmetric states when the length of the spin chain tends to infinity. This representation is dictated by the need to ensure normalization and orthonormality of permutation symmetric states in the thermodynamic limit. In Section \ref{sect6} we compute the bipartite entanglement entropy of certain linear combinations (both finite and infinite) of permutation symmetric states in the thermodynamic limit. In each case, we extract the leading behaviour of the entropy as the size of the block tends to infinity. In Section \ref{sect7} we first perform a similar computation of the bipartite entanglement entropy as in Section \ref{sect6} but we specialize to $s=\frac{1}{2}$ and consider a very special linear combination of permutation symmetric states whose support is the Cantor set projected onto half of a great circle on the Bloch sphere. We then provide a general argument on how the entropy of large blocks may scale for linear combinations whose supports have fractal geometries. Finally, we close the paper with some conclusions and outlook on Section 7.

\sect{Main results and discussion}\label{sectmain}

\subsection{Main results} \label{ssectresults}

Consider a Hilbert space composed of $N$ copies of spin-$s$ representation spaces $\C^{D+1}$, $D=2s$:
\[
	{\cal H}_N = \lt(\C^{D+1}\rt)^{\otimes_\C N}
\]
Let ${\cal P}_N$ be the subspace of ${\cal H}_N$ formed by all vectors that are symmetric under all permutations of sites. It is a simple matter to see that one can describe basis elements for ${\cal P}_N$ by specifying the numbers $N_j$ of sites with $S_z$-eigenvalue equal to $s-j$, $j=0,1,\ldots,D$, under the sole condition $\sum_{j=0}^{D}N_j = N$. Any such basis vector for ${\cal P}_N$ is a suitably normalized linear combination with equal coefficients of all vectors with fixed $N_j$:
\begin{equation}
    |\Psi(N_0, N_1,\ldots,N_{D})\rangle = \sqrt{\frac{N_0! N_1!\ldots N_{D}!}{N!}} \sum_{\sigma \in S_N} |\sigma\Big(\underbrace{\cvec{v}_{0}\cdots \cvec{v}_{0}}_{N_0} \underbrace{\cvec{v}_{1}\cdots \cvec{v}_{1}}_{N_1}
    \ldots \underbrace{{\cvec{v}_{D}\ldots \cvec{v}_{D}}}_{N_{D}}\Big) \rangle, \label{psgs}
\end{equation}
 where the sum is over permutations of the vectors $\cvec{v}_i$, with $i=0,1,\ldots D$. The vectors $\cvec v_i$ are complex $D+1$-dimensional column vectors with one single non-vanishing entry of value 1 at line $i$ (starting from line 0). For every fixed choice of the parameters $N_0,\ldots, N_{D}$ the state (\ref{psgs})
is an eigenstate of the total spin operator $S^z=\sum_{i=1}^N \sigma_i^z$ with eigenvalue $\mu=\sum_{i=0}^{D} (s-i) N_i $. We will refer to the basis vectors (\ref{psgs}) as {\em elementary vectors} (these are also called Dicke states \cite{dicke}).

Consider the particularly simple permutation symmetric vectors where all sites are in the same local state described by $\cvec v$:
\beq\label{psiv}
	|\psi_{\cvec v}\ket = |\cvec v, \cvec v,\ldots,\cvec v\ket.
\eeq
It turns out that it is also possible to write any permutation symmetric vector as a linear combination of $|\psi_{\cvec v}\ket$s. These are more useful for our study of the entanglement entropy, because the vectors (\ref{psiv}) are not entangled, contrary to the elementary vectors (\ref{psgs}). They are the only permutation symmetric vectors that are not entangled. We will refer to (\ref{psiv}) as {\em zero-entanglement vectors}.

In order to see that the zero-entanglement vectors span ${\cal P}_N$,
let us use the following parametrization for normalized complex column vectors,
 \begin{equation}
   \cvec{v}(\underline{\theta},\underline{a}):=\left(
              \begin{array}{c}
                a_0 e^{i \theta_0} \\
                a_1 e^{i\theta_1} \\
                \vdots \\
                a_{D} e^{i\theta_{D}} \\
              \end{array}
            \right) \qquad \text{with} \qquad \sum_{j=0}^{D}a_j^2=1 \quad \text{and}\quad a_j \in \mathbb{R}\quad \forall\, j.\label{con}
 \end{equation}
We use $\underline{\theta}:=(\theta_0,\ldots,\theta_{D})$ with $\theta_i \in [0, 2 \pi]$ for $i=0,\ldots,D$ and $\underline{a}:=(a_0,\ldots,a_{D})$ with $a_i\in [0,1]$ for  $i=0,\ldots,D$. It is a simple matter to see that
\begin{eqnarray}
 |\psi_{\cvec{v}(\underline{\theta},\underline{a})}\rangle=\sqrt{N!}\sum_{N_0,\ldots, N_{D}=0}^N \left[\prod_{j=0}^{D}\frac{a_j^{N_j}}{\sqrt{N_j!}}\right]\delta_{N,\sum_{j=0}^{D}{N_j}} |\Psi(N_0, N_1,\ldots,N_{D})\rangle .\label{es}
\end{eqnarray}
The inversion of (\ref{es}) is obtained by performing the following operations: first, we multiply both sides of the equation by
 $\exp(-i \sum_{j=0}^{D-1} \theta_j \tilde{N}_j)$; finally we integrate both sides of the equation over the variables $\theta_0,\ldots,\theta_{D-1}$, normalizing each integral by a factor of $1/2\pi$. We obtain:
\begin{eqnarray}
&&\frac{1}{(2\pi)^{D}}\int_0^{2\pi} d^D\underline{\theta} \,e^{-i \sum_{j=0}^{D-1} \theta_j \tilde{N}_j}|\psi_{\cvec{v}(\underline{\theta},\underline{a})}\rangle = \frac{\sqrt{N!}}{(2\pi)^{D}}\sum_{N_0,\ldots, N_{D}=0}^N \left[\prod_{j=0}^{D}\frac{a_j^{N_j}}{\sqrt{N_j!}}\right] \nonumber\\
&\times &\int_0^{2\pi} d^D\underline{\theta} \,e^{i \sum_{j=0}^{D-1} \theta_j (N_j-\tilde{N}_j)}e^{i\theta_{D}N_{D}}\delta_{N,\sum_{j=0}^{D}{N_j}}|\Psi(N_0, N_1,\ldots,N_{D})\rangle
\end{eqnarray}
where we employed the shortcut notation:
\begin{equation}\label{dtheta}
    \int_0^{2\pi} d^D\underline{\theta} f(\underline{\theta}):=\int_0^{2\pi} d\theta_0 \int_0^{2\pi} d\theta_1 \ldots \int_0^{2\pi} d\theta_{D-1} f(\theta_0,\ldots, \theta_{D-1}).
\end{equation}
Employing $\delta_{a,b}=\frac{1}{2 \pi}\int_0^{2\pi} e^{i(a-b)\theta} d\theta$ we can carry out the integrals on the r.h.s.~to obtain:
\begin{equation}
  |\Psi(N_0, N_1,\ldots,N_{D})\rangle =
   \frac{1}{\sqrt{N!}}\left(\prod_{j=0}^{D}\frac{\sqrt{N_j!}}{a_j^{N_j}}\right)\frac{1}{(2\pi)^{D}}\int_0^{2\pi} d^D\underline{\theta} \,e^{-i \sum\limits_{j=0}^{D} \theta_j {N}_j}|\psi_{\cvec{v}(\underline{\theta},\underline{a})}\rangle.\label{main}
\end{equation}
We have integrals rather than strictly a linear combination, but it is obvious that these integrals are well defined
(e.g.~as integrals on $\C^{N(D+1)}$) and give a vector in ${\cal P}_N$. It is interesting to note that (\ref{main})
is in fact independent of the choice of the parameters $\underline{a}$. These can in principle be fixed to any particular values,
as long as the normalization condition (\ref{con}) is met. Some choices, however, are more appropriate for certain situations; an  interpretation for these parameters will become apparent later, when the thermodynamic limit is considered. From (\ref{main}) it is easy to show that the state $ |\Psi(N_0, N_1,\ldots,N_{D})\rangle$ still has norm one as it should, using the overlap
\begin{eqnarray}
\langle  \psi_{\cvec{v}(\underline{\hat{\theta}},\underline{a})}|\psi_{\cvec{v}(\underline{\theta},\underline{a})}\rangle= \left(\sum_{j=0}^{D}a_j^2 e^{i(\theta_j-\hat{\theta}_j)} \right)^N,
\label{sp}
\end{eqnarray}
which is immediate from the definitions (\ref{psiv}) and (\ref{con}).

Equation (\ref{main}) means that the zero-entanglement vectors (\ref{psiv}) span ${\cal P}_N$. Clearly, from (\ref{main})
we may as well fix both the length and the overall phase of $\cvec v$ in (\ref{psiv}), and the vectors will still
span ${\cal P}_N$. Hence, we may see $\cvec v$ as homogeneous coordinates for $\CP^D$, and we find that vectors parametrized by points
 in $\CP^D$ span ${\cal P}_N$. This is an over-determination: as we mentioned, it is possible to fix $\underline{a}$, and it would in fact be possible to take only a finite number
  of points in $\CP^D$ in order to span ${\cal P}_N$, although we will not need this here.
  Note that $\CP^{2s}$ is the geometric space of quantum spin-$s$ states,
  studied from this viewpoint in the context of geometric quantum mechanics \cite{gqm1,gqm2,gqm3}. We will discuss this connection below.

Let us now consider the limit $N\to\infty$ of ${\cal P}_N$. This limit of course must be taken with care.
In particular, we need an appropriate topology. Taking the somewhat complicated limit of vectors in the Hilbert
space is not particularly useful for our purposes. Instead, we will consider quantum states seen as linear maps on ${\rm End}({\cal H}_N)$:
\beqa
	\psi\ :\  \End({\cal H}_N) & \to & \C \n
	\Or &\mapsto& \frc{\bra \psi|\Or|\psi\ket}{\bra\psi|\psi\ket},\quad |\psi\ket \in {\cal P}_N
\eeqa
and take limits on these linear maps. We will denote by $\tS_N$ the set of quantum states (linear maps)
corresponding to vectors in ${\cal P}_N$. That is, the set $\tS_N$ is the set of pure permutation symmetric quantum states on $N$ sites.

For quantum states corresponding to vectors in ${\cal H}_N$ in general, instead of ${\cal P}_N$, one would require, in order to define
the large-$N$ limit, an embedding of ${\cal H}_N$ into ${\cal H}_{N+1}$ for every $N$: a prescription
as to where the sites are being added (in more mathematical terms, in order to define the limit set of operators, one would have to construct a direct system on the algebras of linear operators on ${\cal H}_N$, and take the direct, or inductive, limit). However, since we are looking at permutation symmetric states, there is
 a canonical embedding. This goes as follows. Any operator $\Or\in {\rm End}({\cal H}_N)$ is a linear combination
 of products of operators ${\cal A}=A_1\cdots A_N$ factorized on the tensor factors of ${\cal H}_N$. By permutation
  symmetry, when evaluating $\psi(A)$ with $\psi\in\tS_N$, the factors acting nontrivially (i.e.~different from the identity) can always be gathered contingently
  on the sites, say, at positions $1,2,3,\ldots$, and in any order. Hence, for a factorized operator ${\cal A}$, the
   only information that we need in order to evaluate $\psi({\cal A})$ is the set of
   nontrivial factors, as matrices on $\C^{D+1}$ (we do not need the information of the site on which they act).
   The same information can be provided to evaluate $\psi'({\cal A})$ for any $|\psi'\ket \in {\cal P}_M$, as long as the number
    of nontrivial factors is less than or equal to $M$. This, along with linearity, gives the canonical embedding if $M>N$.

Let us refer to any linear operator $\Or\in\End({\cal H}_N)$, for any finite $N=1,2,3,\ldots$, as {\em finitely-supported} (following the terminology of \cite{permutation}).
Let $\psi_N\in\tS_N$, $N=1,2,3,\ldots$ be an infinite sequence of quantum states.
 We will say that $\lim_{N\to\infty} \psi_N$ exists in the {\em local-operator topology} if $\lim_{N\to\infty} \psi_N(\Or)$ exists for every
 finitely-supported $\Or$ (with the embedding described above). This defines a linear map on the space of finitely-supported operators,
  $\psi=\lim_{N\to\infty}\psi_N$ with $\psi(\Or) = \lim_{N\to\infty} \psi_N(\Or)$. Let us denote by $\tS$ the space of all such linear maps, that occur as limits in the local-operator topology.
   We may see this as the space of all permutation symmetric quantum states in the thermodynamic limit\footnote{The set of limits $\lim_{N\to\infty}\psi_N$, with $\psi_N$ sequences of pure permutation symmetric quantum states, is expected to give rise, from the viewpoint of local operators, to all thermodynamic permutation symmetric quantum states, pure or mixed; but we do not have a proof of this.}. As described above, if $\psi\in\tS$
   and ${\cal A}$ is a finitely supported operator that is factorized, then $\psi({\cal A})$ only depends on the non-trivial factors of ${\cal A}$.
   Also, $\psi$ has the properties that $\psi({\bf 1}) = 1$ and that $\psi(\Or)^*=\psi(\Or^\dag)$, as usual for quantum states. Note that an explicit metric inducing the local-operator topology
   (in a slightly more general context) was given in \cite{permutation}.

For any fixed $\cvec v$, the $N\to\infty$ limit of the sequence of quantum states corresponding to zero-entanglement vectors $|\psi_{\cvec v}\ket$ defined in (\ref{psiv}) is obviously convergent in the local-operator topology. By a slight abuse of notation, we will denote it by $\psi_{\cvec v}\in \tS$, and refer to it as a zero-entanglement state. Our first result is as follows.
\begin{theorem} \label{th1}
Let $\psi\in \tS$. Then there exists a probability measure $\mu$ on $\CP^D$,
\[
	\int_{\CP^D} d\mu(\cvec v) = 1,
\]
such that
\beq\label{psiO}
	\psi(\Or) = \int_{\CP^D} d\mu(\cvec v)\, \psi_{\cvec v}(\Or)
\eeq
for every finitely-supported $\Or$. Further, every probability measure gives rise to a unique $\psi\in \tS$.
\end{theorem}
Note that we expect the measure to be unique, since in particular $\CP^D$ is compact (there is obviously unicity up to the weak equivalence under equality of all averages of finitely-supported operators).

Basically, what is happening is that all of $\CP^D$ becomes necessary to span ${\cal P}_N$ in the large-$N$ limit.
In fact, $\CP^D$ becomes a set of basis elements, and further $\bra\psi_{\cvec v}|\Or|\psi_{{\cvec v}'}\ket \to 0$
whenever $\cvec v$ and $\cvec v'$ are not colinear, for every finitely-supported operator $\Or$. Hence, any linear
combination $|\psi_N\ket = \sum_{\cvec v} c_{\cvec v} |\psi_{\cvec v}\ket$ gives rise, in the large-$N$ limit,
 to the quantum state $\sum_{\cvec v} |c_{\cvec v}|^2 \psi_{\cvec v}$ formed purely out of $\psi_{\cvec v}$,
  because the cross-terms vanish. The sum may contain infinitely many terms,
  and the more precise way of describing this is using a measure: $\sum_{\cvec v} |c_{\cvec v}|^2 \mapsto \int _{\CP^D} d\mu (\cvec v)$.

In the present paper we will not present a complete proof of Theorem \ref{th1}, but in Section \ref{sectThermo} we will give various derivations which provide a good intuition for (\ref{psiO}). The complete proof of Theorem \ref{th1} will be presented in a forthcoming work.

The quantity of interest to us is the entanglement entropy (\ref{renyi}). Our next result shows that the above formalism is useful
to evaluate the entanglement entropy in the thermodynamic limit.
\begin{theorem}\label{th2}
Let $\psi_N\in\tS_N$, $N=1,2,3,\ldots$ be a sequence of quantum states converging as $N\to\infty$ to $\psi\in \tS$ in the local
operator topology. Then for every integer $n>1$, the R\'enyi entropy of the states $\psi_N$, with a fixed number $m$ of sites in a subsystem of the bipartition, converges as $N\to\infty$ to
\beq\label{Sn}
	S_n = \frc{1}{1-n} \log\lt[\int_{(\CP^D)^{\times n}}
	\lt(\prod_{\alpha=1}^n d\mu(\cvec v_\alpha)\rt)
	\lt(\prod_{\alpha=1}^n \frc{\cvec v_\alpha^{\, \dag} \cdot\cvec v_{\alpha+1}}{\cvec v_\alpha^{\, \dag}\cdot\cvec v_\alpha}\rt)^m\rt]
\eeq
where $\mu$ is the measure of Theorem \ref{th1}, and $\cvec v_\alpha^{\, \dag}$ is, as usual, the transpose of
the complex conjugate of the column vector $\cvec v_\alpha$. Here we use $\cvec v_{n+1}:=\cvec v_1$.
\end{theorem}
Note that on the right-hand side of (\ref{Sn}) the integrand indeed lies on $(\CP^D)^{\times n}$;
 in particular, the phases of $\cvec v_\alpha$ are unimportant thanks to cyclicity of the product. We provide a proof of Theorem \ref{th2} in Section \ref{sectth2}

There is a canonical metric on $\CP^D$: one considers the metric on the sphere
 $S^{2D+1} = \{(z_0,z_1,\ldots,z_D)\in \C^{D+1}:|z_0|^2 + \ldots + |z_{D}|^2=1\}$,
 and takes for $\CP^D = S^{2D+1}/S^1$ the quotient metric induced by simultaneous phase shifts
 on all coordinates $z_i$. This is the Study-Fubini metric \cite{fubini,study,sfbook1,sfbook2}
 (also referred to as the Bures-Uhlmann metric in the context of geometric
  quantum mechanics \cite{bures,uhlmann}), giving in particular a differentiable manifold
   structure to $\CP^D$ of real dimension $2D$. Our next results, which are the
    main results of this paper, show that the entanglement entropy is naturally associated to this manifold structure and in particular to this metric.

\begin{conjecture}\label{th3}{\ }\\
{\bf I}. Let the measure $\mu$ of Theorem \ref{th1} be supported on a subset $U$ of $\CP^D$ with (possibly fractal) dimension $d$ with respect to the Study-Fubini (Lipshitz-)manifold structure. Then the large-$m$ behaviour of $S_n$ is given by
\beq\label{Snass1}
	S_n \sim \frc{d}2 \log m + O(1)
\eeq
for every real $n\geq1$. Note that $0\leq d\leq 2D=4s$, and all cases are possible.\\
{\bf II}. Assume that the subset $U$ where $\mu$ is supported is a submanifold of $\CP^D$. Assume further that $\mu$ is absolutely
continuous with respect to the volume element on $U$ induced by the Study-Fubini metric on $\CP^D$. That is, there exists a density
 $f(\cvec v)$ such that $d\mu(\cvec v) = d^d\cvec v f(\cvec v)$ where $d^d\cvec v$ is the volume element on $U$. Then\footnote{Note that for
 spin-$\frac{1}{2}$ this result
 was first given in \cite{fractal}, albeit containing some typos.}
\beq\label{Snass2}
	S_n \sim \frc{d}2 \log \frc{m}{8\pi} + \frc1{1-n}\log\lt(n^{-\frc d2}
	\int d^d\cvec v\, f(\cvec v)^n\rt) + o(1).
\eeq
\end{conjecture}
In Part I of this conjecture, we expect only the Lipshitz-manifold structure to be important, as the fractal dimension is invariant under bi-Lipshtiz transformations.

Results (\ref{Snass1}) and (\ref{Snass2}) are derived from (\ref{Sn}) taking $n>1$ integer. However, they make
sense as well for real $n\geq 1$. We conjecture that they are the correct asymptotic results for the R\'enyi entropy
 for all real $n\geq 1$. A general calculation as well as various examples for Part II of Conjecture \ref{th3} are provided in Section \ref{sect6}, and for Part I, in Section \ref{sect7}.
 
 With this conjecture, we find in particular for the von Neumann entanglement entropy,
\beq\label{Snass3}
	S_1\sim \frc{d}2 \log \frc{em}{8\pi} - \int d^d\cvec v\,f(\cvec v)\log f(\cvec v) + o(1)
\eeq
We observe that the second term in this expression is the so-called geometric entropy of geometric quantum mechanics.
 This is the entropy associated to the probability density $f(\cvec v)$. 
 
Another interesting limit is the limit $n\to\infty$, which gives the single-copy entropy $S_\infty = -\log \lambda_1$ where $\lambda_1$ is the largest eigenvalue of the reduced density matrix. A simple saddle-point analysis gives, in the case where $f_{\rm max} := {\rm max}_{\cvec v}f( \cvec v) <\infty$,
\beq
	S_\infty \sim \frc{d}2 \log \frc{m}{8\pi} - \log(f_{\rm max}) + o(1).
\eeq

Note that in all cases, the leading large-$m$ behaviour is the same, in contrast to the results for critical one-dimensional quantum chains.

Note that in (\ref{Snass2}) and (\ref{Snass3}), the Study-Fubini metric is singled out: it is the density $f(\cvec v)$ with respect to the Study-Fubini-induced volume element that occurs, and a different choice
of density would lead to different-looking, more complicated formulae, since $f(\cvec v)^n$ and $f(\cvec v)\log f(\cvec v)$ are not densities (in particular, there would be $n$-dependent extra factors under the integral).

To be more precise, the normalization of the metric that we choose is such that, in particular, the sphere $\CP^1$ has radius 1.
The distance function is, in homogeneous coordinates \cite{gqm3,hc},
\beq\label{distCP}
	D(\cvec v,\cvec w) = 2\,{\rm arccos}\,\sqrt{\frc{
	\cvec v^{\,\dag} \cvec w\;\cvec w^{\,\dag} \cvec v}{
	\cvec v^{\,\dag} \cvec v\;\cvec w^{\,\dag} \cvec w}}.
\eeq
We can write down the Study-Fubini metric and volume element explicitly. Consider the coordinates $\theta_i\in[0,2\pi)$, $i=0,\ldots,D-1$ as well as the spherical coordinates $\phi_i$, $i=0,\ldots,D-1$ on the sphere $\{(a_0,a_1,\ldots,a_D)\in \R^{D+1}:\sum_{i=0}^{D} a_i^2=1\}$,
    \begin{equation}\label{spherical}
      a_0= \cos \phi_0,\quad  a_i=\cos\phi_i \prod_{k=0}^{i-1}\sin \phi_k\quad \text{for} \quad i=1,\ldots,D-1\quad\text{and}\quad a_{D}=  \prod_{k=0}^{D-1}\sin \phi_k.
    \end{equation}
The $a_i$ are nonnegative, hence
    \begin{equation}\label{intervals}
        0 \leq \phi_i \leq \frac{\pi}{2}\quad \text{for}\quad i=0,\ldots,D-1.
    \end{equation}
In terms of this parametrization, the Study-Fubini metric is
\beq\label{metric}
	g_{\theta_k,\theta_j} = -4(a_k^2a_j^2 - a_k^2\delta_{k,j}),\quad
	g_{\phi_k,\phi_j} = 4\delta_{k,j}\,\prod_{p=0}^{k-1}\sin^2\phi_p
\eeq
leading in particular to the volume element
\beq\label{volelem}
	d^{2D}\cvec v = 2^D\prod_{p=0}^{D-1} \sin2\phi_p\;
	\prod_{k=0}^{D-1}\prod_{p=0}^{k-1} \sin^2\phi_p\;
	d^D\underline\theta \,d^D\underline\phi\;.
\eeq

\subsection{Discussion}

{\bf Geometric quantum mechanics}. Our results have connections to geometric quantum mechanics. Indeed,
in that context, one also uses probability measures $\mu$ on $\CP^D$, in order to describe the mixed states
 of statistical quantum mechanics. As is well known, this description is rather redundant from the viewpoint
 of linear quantum mechanics: one can always diagonalize the density matrix, hence choose a measure
 supported on the points corresponding to the eigenvectors. An important difference, in our set-up, is that there is no
  redundancy: different measures $\mu$ correspond to different thermodynamic permutation symmetric quantum states.

An natural quantity in the context of geometric quantum mechanics is the geometric entropy $-\int d^d \cvec v \, f(\cvec v)\log f(\cvec v)$ associated to a Study-Fubini density $f(\cvec v)$. This quantity, in particular, partially lifts the redundancy mentioned above, hence its meaning, in the context of linear quantum mechanics, is not always clear. Further, one can ask about the maximization of the geometric entropy under the condition
 of a fixed averaged energy, controlled by the temperature $T$. It turns out that it does not lead to the usual thermal density matrix of statistical mechanics,
 where eigenstates of the energy appear with Maxwell-Boltzmann probabilities. Rather \cite{gqm3}, it leads to the density
\beq
	f_*(\cvec v) = \frc{e^{-H(\cvec v)/T}}{\int_{\CP^D} d^{2D}\cvec v\,
	e^{-H(\cvec v)/T}},
\eeq
where $H(\cvec v)$ is the average of the energy in the quantum state pointed by the homogeneous coordinates
$\cvec v$ of $\CP^D$. It can be shown that this is not simply a redundant expression for the standard thermal
 density matrix; it gives a different linear mixed state \cite{gqm3}. Our set-up provides an appealing physical interpretation
 for $f_*$. It is the distribution for thermodynamic quantum states obtained by maximizing the entanglement entropy of a large subsystem, under the condition of a fixed average energy,
  and under the extra condition that the system be restricted to the subset of permutation symmetric states.

\vspace{0.2cm}

\noindent {\bf Creating entanglement.} A standard way of creating entanglement in the context of quantum information
 theory is by symmetrizing a vector formed by factorized qubits. The resulting states are usually referred to as Dicke states \cite{dicke}.
 Our results (\ref{Snass1}), (\ref{Snass2}) provide an upper bound for the entanglement that can be stored by this method in a large subsystem
 of $m$ spin-$s$ qubits singled out from a much larger system of spin-$s$ qubits. The entanglement can grow logarithmically (as already found
 by Popkov et al. \cite{pop2}), with a coefficient that can be as large as $D=2s$,
 but not larger. Further, when this bound is achieved, the entanglement is maximized by maximizing the geometric entropy associated to the coefficients obtained by expanding in the zero-entanglement vectors (\ref{psiv}).
 
 Note that in \cite{pop2} it was stated incorrectly that the logarithmic growth for the entanglement of pure permutation symmetric states couldn't have a coefficient larger than $s$. It was however remarked that for mixed states, the coefficient could be as large as $2s$. Here we have not analyzed mixed states, but we expect that the set $\tS$, although obtained as thermodynamic limits of pure permutation symmetric states only, contains also all mixed permutation symmetric states; whence our results being consistent with Popkov et al.'s remark. This is based on the known fact that any mixed state on $N$ sites can be reproduced using a pure state on $2N$ sites. Hence in the large-$N$ limit, from the viewpoint of finitely-supported operators, there is no distinction between mixed and pure states. This argument is much more subtle when one restricts to permutation symmetric states, because the doubling procedure of a mixed permutation symmetric state on $N$ sites does not, in general, produce a pure permutation symmetric state on $2N$ sites. It would be interesting to analyze this situation in more depth.

\vspace{0.2cm}

\noindent {\bf Comparison with previous results.} In \cite{pop2}, the limit $N\to\infty$ was taken on the entanglement entropy (in fact, on the reduced density matrix) associated to the elementary vectors $|\Psi(N_0,N_1,\ldots, N_D)\ket$ by keeping fixed ratios $N_j/N = p_j$ (more precisely, by ``keeping fixed ratios $N_j/N = p_j$'' we will mean taking $N_j= N_j(N)$ such that $\lim_{N\to\infty}N_j(N)/N =  p_j$). Clearly, we have $\sum_{j=0}^D p_j = 1$. We will show that this limit exists in the local operator topology on the elementary quantum states $\Psi(N_0,N_1,\ldots, N_D)$ themselves:
\beq\label{popovstates}
	\Psi(p_0,p_1,\ldots,p_D) := \lim_{N\to\infty} \Psi(N_0,N_1,\ldots, N_D).
\eeq
We will show that in this case, the measure $\mu$ above is supported on the submanifold specified by $a_i = \sqrt{p_i}$, $i=0,1,\ldots,D$, and that on this submanifold, we have
\beq
	d\mu(\cvec{v}(\underline\theta,\underline a))
	= \frc1{(2\pi)^D} d\theta_0\cdots d\theta_{D-1},\quad \theta_D = 0.\label{sub}
\eeq
We provide the derivations (but no mathematically rigorous proofs) in Subsection \ref{ssectthermopopov}.

This provides the interpretation of the coefficients $a_i$ that were arbitrary in (\ref{main}): they must be fixed to the square-roots of the ratios $N_j/N$ in order for the thermodynamic limit to make sense on (\ref{main}).
In particular, the large-$N$ limit considered in \cite{pop2} is not the most general one. The leading logarithmic divergence found in \cite{pop2} was $s\log m$, because the chosen submanifold is of dimension $d=D=2s$, half of the real dimension of the $\CP^D$ manifold.

For more precision, let us note that the metric induced on the submanifold $a_i = \sqrt{p_i}$ is $g_{\theta_j,\theta_k}$ (\ref{metric}). It has volume element
\[
	d^D\cvec v = 2^D\prod_{i=0}^{D} a_i d^D\underline{\theta}.
\]
Hence, we see that the conditions of Conjecture \ref{th3}.II are satisfied, with
\[
	f = \frc1{(4\pi)^D \prod_{i=0}^D a_i}.
\]
Therefore, equation (\ref{Snass2}) gives us, in terms of $s=D/2$ and $p_i=a_i^2$,
\begin{equation}\label{renyi2}
S_n \sim s\log 2\pi m + \frc{s\log n}{n-1} + \frc12 \log \prod_{i=0}^{D} p_i + o(1).
\end{equation}
For $n=1$ we recover the result obtained in \cite{pop2}, namely the large $m$ behaviour of the von Neumann entropy:
\begin{equation}
    S_1\sim s\log 2\pi e m  + \frc12 \log \prod_{i=0}^{D} p_i + o(1).
    \label{vn3}
\end{equation}

Finally, let us note that for spin $1/2$, it was found in \cite{vidal} that the ground state of the Lipkin-Meshkov-Glick model, for certain values of the parameters where it is not a pure Dicke state, has an entanglement entropy with the logarithmic behaviour (\ref{ent}) with $d=2/3$. It would be very interesting to understand this state from a geometric point of view.

\sect{Proof of Theorem \ref{th2}} \label{sectth2}

A proof of this result for $s=\frac{1}{2}$ was sketched in the letter \cite{fractal}. Here we will present an extended and more general proof.
The starting point is the expression of the entropy in terms of local cyclic replica permutation operators $\mathcal{T}_i$ which act on site $i$ of a
quantum spin chain and cyclicly permute the spins of $n$ replicas of the model at that same site $i$. These operators were introduced in  \cite{permutation}
where we also showed that
the R\'enyi entropy of a block of (not necessarily consecutive) spins of size $m$ is given by
\begin{equation}\label{tt}
    S_n=\frac{1}{1-n} \log \left[\frac{\langle\psi^{\otimes n}| \mathcal{T}_A|\psi^{\otimes n}\rangle}{\langle\psi^{\otimes n}|\psi^{\otimes n}\rangle}\right],
\end{equation}
where $\psi^{\otimes n}$ is the $n$-th tensor power of the state $\psi$ and
\begin{equation}
   \mathcal{T}_A = \prod_{i \in A} \mathcal{T}_i\quad \text{and} \quad |A|=m.
\end{equation}
Employing the language of functionals as in the previous section, we can express the correlation function above simply as
\begin{equation}
    S_n=\frac{\log\left(\psi^{\otimes n}(\mathcal{T}_A)\right)}{1-n},
    \label{here}
\end{equation}
From \cite{permutation} we also know that
\begin{equation}
    \mathcal{T}_i=\text{Tr}_{\text{aux}}\left(\prod_{\alpha=1}^n \sum_{\epsilon_1,\epsilon_2=1}^{D+1} E^{\text{aux}}_{\epsilon_1 \epsilon_2}
    E^{\alpha,i}_{\epsilon_2 \epsilon_1}\right),
\end{equation}
where $E_{\epsilon_{1}\epsilon_2}$ are the $(D+1) \times (D+1)$ elementary matrices, where all elements are zero except for a 1 in row $\ep_1$, column $\ep_2$, and the indices $\text{aux}$, $\alpha$ and $i$ refer to
an auxiliary space $V_\text{aux}=\mathbb{C}^{D+1}$, the copy number and the site, respectively. Since $\mathcal{T}_A$, for every finite set of sites $A$, is a finitely-supported operator, and since the entanglement entropy is expressed as the evaluation of this operator in a quantum state, then by Theorem \ref{th1} the thermodynamic limit of the entanglement entropy exists and can be expressed using (\ref{psiO}).

We can now write the R\'enyi entropy of the
infinite chain by recalling (\ref{psiO}) and applying it to the replica case and the operator $\mathcal{T}_A$
\beq\label{psiO2}
	\psi^{\otimes n}(\mathcal{T}_A) = \int_{(\CP^D)^{\times n}} \left(\prod_{\alpha=1}^n d\mu(\cvec v_\alpha)\right)\, \bigotimes_{\alpha=1}^n\psi_{\cvec v_\alpha}(\mathcal{T}_A),
\eeq
with
\begin{equation}
   \bigotimes_{\alpha=1}^n\psi_{\cvec v_\alpha}(\mathcal{T}_A)=\prod_{i\in A}
   \text{Tr}_{\text{aux}}\left(\prod_{\alpha=1}^n \sum_{\epsilon_1,\epsilon_2=1}^{D+1} E^{\text{aux}}_{\epsilon_1 \epsilon_2}
    \psi_{\cvec v_\alpha,i}(E^{\alpha,i}_{\epsilon_2 \epsilon_1})\right),\label{mess}
\end{equation}
where we used, in an obvious notation,
\begin{equation}\label{tpr}
   \psi_{\cvec v_\alpha}=\bigotimes_{i \in \mathbb{Z}} \psi_{\cvec v_\alpha,i}.
\end{equation}
The quantity $ \psi_{\cvec v_\alpha,i}(E^{\alpha,i}_{\epsilon_2 \epsilon_1})$ is independent of the site $i$, and from (\ref{psiv}) and (\ref{tpr}),
 \begin{equation}
    \psi_{\cvec v_\alpha,i}(E^{\alpha,i}_{\epsilon_2 \epsilon_1})= v_{\alpha,\epsilon_2}^* v_{\alpha,\epsilon_1}.
 \end{equation}
 Substituting into (\ref{mess}) and tracing over the auxiliary space we find
\beqa
 \text{Tr}_{\text{aux}}\left(\prod_{\alpha=1}^n \sum_{\epsilon_1,\epsilon_2=1}^{D+1} E^{\text{aux}}_{\epsilon_1 \epsilon_2}
    v_{\alpha,\epsilon_2}^* v_{\alpha,\epsilon_1}\right)&=&
    \sum_{\ep_1,\ldots,\ep_n=1}^{D+1}
    v_{1,\ep_1}v^*_{1,\ep_2} v_{2,\ep_2}v^*_{2,\ep_3}\cdots
    v_{n,\ep_n}v^*_{n,\ep_1} \n
    &=&
    \prod_{\alpha=1}^n \cvec{v}_\alpha^\dagger\cdot \cvec{v}_{\alpha+1}\,.
\eeqa
 Therefore
 \beq\label{psiO3}
	\psi^{\otimes n}(\mathcal{T}_A) = \int_{(\CP^D)^{\times n}} \left(\prod_{\alpha=1}^n d\mu(\cvec v_\alpha)\right)\, \left(\prod_{\alpha=1}^n
\cvec{v}_\alpha^\dagger\cdot \cvec{v}_{\alpha+1}\right)^m.
\eeq
 which when substituted in (\ref{here}) gives the result (\ref{Sn}) for $\cvec{v}_\alpha^\dagger \cdot \cvec v_\alpha=1$.

\sect{Thermodynamic limit of permutation symmetric states}\label{sectThermo}

The complete proof of Theorem \ref{th1} is rather involved, and slightly beyond the scope of the present paper. We instead present, here, a proof of (\ref{psiO}) in the case of finite linear combinations of vectors $|\psi_{\cvec v}\ket$ in fixed directions $\cvec v$, and a derivation (not of mathematical rigor) of the expression (\ref{psiO}), (\ref{sub}) for the states (\ref{popovstates}). The latter are the thermodynamic limit $N\to\infty$ of the states $\Psi(N_0,N_1,\ldots,N_D)$ associated to (\ref{psgs}), with the condition that the ratios $N_j/N = p_j$ be fixed. This not only will give us some intuition as to the way the measure $\mu$ of Theorem \ref{th1} appears, but also will be immediately relevant to the analysis, within our framework, of the results of Popov et al. \cite{pop2}. We will also consider the cases of finite linear combinations of the vectors $|\Psi(N_0,N_1,\ldots,N_D)\ket$, held in the thermodynamic limit at different sets of ratios $\{p_j\}$.

\subsection{Finite linear combinations}

Let $V$ be a finite set of complex vectors $\cvec v$ of unit length $\cvec v^{\,\dag}\cvec v = 1$, such that no two of them are colinear (i.e.~related to each other by an overall phase). Let us consider $|\psi_N\ket = \sum_{\cvec v\in V}{c_{\cvec v,N}} |\psi_{\cvec v}\ket\in{\cal H}_N$, with the condition that $\lim_{N\to\infty} c_{\cvec v,N} = c_{\cvec v}$ exists. For every finitely-supported operator $\Or$, which acts nontrivially on $\ell$ sites, we have, for all $N\geq \ell$,
\beq
	\bra\psi_{\cvec v}|\Or|\psi_{\cvec w}\ket =
	\bra\psi_{\cvec v}|\psi_{\cvec w}\ket \chi_{\cvec v,\cvec w}(\Or)
\eeq
where the factor
\beq\label{chivw}
	\chi_{\cvec v,\cvec w}(\Or) = \lt(\cvec v^{\,\dag}\rt)^{\otimes \ell}
	\Or\, \cvec w^{\,\otimes \ell}
\eeq
is independent of $N$. In particular, we note that $\chi_{\cvec v,\cvec w}(\Or)$ is a polynomial in the components of $\cvec v$ and $\cvec w$ (hence continuous), and that
\beq\label{chip}
	\chi_{\cvec v,\cvec v}(\Or) = \psi_{\cvec v}(\Or)
\eeq
(recall that by our convention, $\psi_{\cvec v}$ is the limit $N\to\infty$ of the quantum states associated with the finite-$N$ vectors $|\psi_{\cvec v}\ket$). Clearly, we have $\bra\psi_{\cvec v}|\psi_{\cvec w}\ket = (\cvec v^{\,\dag}\cvec w)^N$. The maximal value of $|\cvec v^{\,\dag}\cvec w|$ occurs when $\cvec v$ and $\cvec w$ are colinear, hence equal to each other if both are in $V$. This maximal value is 1. Hence,
\beq
	\lim_{N\to\infty} \bra\psi_{\cvec v}|\psi_{\cvec w}\ket
	= \lt\{\ba{ll} 1 & (\cvec v = \cvec w) \\ 0 & (\cvec v \neq \cvec w) \ea\rt.
\eeq
whenever $\cvec v,\cvec w\in V$. This yields
\beqa
	\lim_{N\to\infty}\bra\psi_N|\Or|\psi_N\ket &=&
	\lim_{N\to\infty}
	\sum_{\cvec v,\cvec w\in V}c_{\cvec v,N}^* c_{\cvec w,N}
	\bra\psi_{\cvec v}|\psi_{\cvec w}\ket \chi_{\cvec v,\cvec w}(\Or) \n
	&=&	\sum_{\cvec v\in V} |c_{\cvec v}|^2
	\psi_{\cvec v}(\Or)
\eeqa
which shows (\ref{psiO}) in the cases of finite linear combinations, with
\beq\label{mufinite}
	\int d\mu(\cvec v)\,F(\cvec v) = \sum_{\cvec v\in V} |c_{\cvec v}|^2\,F(\cvec v).
\eeq
That is, the measure is supported on the finite set $V$, with weights $|c_{\cvec v}|^2$:
\beq
	\lim_{N\to\infty} \psi_N = \sum_{\cvec v\in V} |c_{\cvec v}|^2
	\psi_{\cvec v}.
\eeq

\subsection{Elementary states}\label{ssectthermopopov}

We now consider the thermodynamic limit of the elementary states $\Psi(N_0,N_1,\ldots,N_D)$ associated to the elementary vectors (\ref{psgs}). In order to apply a similar methodology, we  use the expression (\ref{main}). As we said, the ratios $\frac{N_i}{N}=p_i$ remain finite, and since $N=\sum_{i=0}^{D}N_i$ it follows that $\sum_{i=0}^{D} p_i=1$. For simplicity of the derivation, we will assume that $p_j\neq0$ for all $j$, but this is not essential and the final result holds for general $p_j$.

The intuition behind (\ref{psiO}) is that, as is clear form the previous subsection, the vectors $|\psi_{\cvec v}\ket$  become orthogonal, at different values of $\cvec v$, in the limit $N\to\infty$. Since in this limit all directions of $\cvec v$ are independent, it is natural to expect that in regular enough situations, like that of $\Psi(N_0,N_1,\ldots,N_D)$, we should normalize the resulting vectors with a delta-function normalization. If, for instance, only the angles $\theta_j$ may vary, then we will have vectors $|\Psi_{\cvec{v}(\underline{\theta},\underline{a})}\rangle$ representing the thermodynamic limit of the vectors $|\psi_{\cvec{v}(\underline{\theta},\underline{a})}\rangle$ taken in such a way as to ensure that they form a basis of factorizable, orthonormal states under the following delta-function normalization
\begin{equation}\label{nop}
  \langle \Psi_{\cvec{v}(\underline{\hat{\theta}},\underline{a})} |\Psi_{\cvec{v}(\underline{\theta},\underline{a})}\rangle= \prod_{j=0}^{D-1}{{2\pi}}\delta(\theta_j-{\hat{\theta}}_j).
\end{equation}
The vector $|\Psi_{\cvec{v}(\underline{\theta},\underline{a})}\rangle$ is, formally, a limit of the form $\lim_{N\to\infty} R_{\cvec v}(N)|\psi_{\cvec v}\ket$ for an appropriate renormalization factor $R_{\cvec v}(N)$. Note that (\ref{nop}) implies, for every finitely supported operator $\Or$,
\beqa
	\bra\Psi_{\cvec{v}(\underline{\hat{\theta}},\underline{a})}|
	\Or |\Psi_{\cvec{v}(\underline{\theta},\underline{a})}\rangle &=&
	\lim_{N\to\infty}
	R_{\cvec v(\underline{\hat{\theta}},\underline{a})}(N)^*
	R_{\cvec v(\underline{{\theta}},\underline{a})}(N)\,
	\bra\psi_{{\cvec v}(\underline{\hat{\theta}},\underline{a})}|
	\Or|\psi_{{\cvec v}(\underline{{\theta}},\underline{a})}\ket \n
	 &=&
	\lim_{N\to\infty}
	R_{\cvec v(\underline{\hat{\theta}},\underline{a})}(N)^*
	R_{\cvec v(\underline{{\theta}},\underline{a})}(N)\,
	\bra\psi_{{\cvec v}(\underline{\hat{\theta}},\underline{a})}|
	\psi_{{\cvec v}(\underline{{\theta}},\underline{a})}\ket\,
	\chi_{\underline{\hat{\theta}},\underline{\theta}}(\Or) \n
	 &=&
 	\bra\Psi_{\cvec{v}(\underline{\hat{\theta}},\underline{a})}|
	\Psi_{\cvec{v}(\underline{\theta},\underline{a})}\rangle\,
	\chi_{\underline{\hat{\theta}},\underline{\theta}}(\Or) \n
	 &=&
 	\psi_{\cvec v(\underline{{\theta}},\underline{a})}(\Or)
	\prod_{j=0}^{D-1}{{2\pi}}\delta(\theta_j-{\hat{\theta}}_j).
	\label{pop}
\eeqa
Here, the factor $\chi_{\underline{\hat{\theta}},\underline{\theta}}(\Or):=\chi_{\cvec{v}(\underline{\h\theta},\underline{a}),\cvec{v}(\underline{\theta},\underline{a})}(\Or)$ (see (\ref{chivw})) is independent of $N$, a fact which we used on the third line in order to evaluate the limit. On the last line, we used (\ref{chip}).

It turns out that in the thermodynamic limit we can express (formally) the vector
\begin{equation}\label{limit}
|\Psi(p_0, p_1,\ldots,p_{D})\rangle:=\lim_{N,N_0,\ldots,N_{D}\rightarrow \infty}
|\Psi(N_0, N_1,\ldots,N_{D})\rangle
\end{equation}
as
\begin{eqnarray}
  |\Psi(p_0, p_1,\ldots,p_{D})\rangle =\frac{1}{(2\pi)^{D}}
    \int_0^{2\pi} d^D \underline{\theta} \,|\Psi_{\cvec{v}(\underline{{\theta}},\underline{{a}})}\rangle,\label{main2}
\end{eqnarray}
where the integration is on $\theta_0,\ldots,\theta_{D-1}$ with $\theta_D=0$ fixed, and the $a_j$ are fixed in terms of the $p_j$ as
\beq\label{ap}
	a_j = \sqrt{p_j}.
\eeq
This implies the expression (\ref{psiO}), with integration on the submanifold given by the fixed $a_j$ and measure $d^D\underline{\theta}/(2\pi)^D$ as claimed in (\ref{sub}). Indeed,
\beqa
	\Psi(p_0, p_1,\ldots,p_{D})(\Or)
	&=& \bra \Psi(p_0, p_1,\ldots,p_{D})|\Or
	|\Psi(p_0, p_1,\ldots,p_{D})\ket \n
	&=& \frc1{(2\pi)^{2D}}
	\int_0^{2\pi} d^D \underline{\h\theta}
	\int_0^{2\pi} d^D \underline{\theta}\,
	\bra\Psi_{\cvec{v}(\underline{\h{\theta}},\underline{{a}})}|
	\Or |\Psi_{\cvec{v}(\underline{{\theta}},\underline{{a}})}\ket \n
	&=& \frc1{(2\pi)^{D}}
	\int_0^{2\pi} d^D \underline{\theta}\,
	\psi_{\cvec v(\underline{{\theta}},\underline{a})}(\Or)
\eeqa
where we used (\ref{pop}).

We now present a derivation to find the coefficient $R_{\cvec v}(N)$ such that (\ref{nop}) holds, and to show that (\ref{main2}) holds with (\ref{ap}).

Let us start by considering the numerical coefficient in (\ref{main})
\begin{equation}
   \frac{1}{{\sqrt{N!}}}\left(\prod_{j=0}^{D}\frac{{\sqrt{N_j!}}}{a_j^{N_j}}\right)\sim (\sqrt{2\pi N})^{D}
   \prod_{j=0}^{D}{p_j}^{\frac{1}{4}}\left(\frac{ \sqrt{p_j}}{a_j}\right)^{N_j},\label{st}
\end{equation}
where we have used Stirling's formula $A!\approx \sqrt{2\pi A} \frac{A^A}{e^A}$ for $A$ large. We can deduce from (\ref{main}), (\ref{sp}) and (\ref{st}) what the state $|\Psi_{\cvec{v}(\underline{\theta},\underline{a})}\rangle$ must be. Since $|\Psi(p_0, p_1,\ldots,p_{D})\rangle$ must have norm one, it follows that
\begin{eqnarray}
1 = \frac{N^{D}}{(2\pi)^{D}}
   \prod_{j=0}^{D}{p_j}^{\frac{1}{2}}\left(\frac{ \sqrt{p_j}}{a_j}\right)^{2N_j}
\int_0^{2\pi} d^D \underline{\theta}\int_0^{2\pi} d^D \underline{\hat{\theta}}\,
 e^{-i \sum_{j=0}^{D} N_j (\theta_j-\hat{\theta}_j)}\left(\sum_{j=0}^{D}a_j^2 e^{i(\theta_j-\hat{\theta}_j)} \right)^N,
\end{eqnarray}
which may be rewritten as
\begin{eqnarray}
 1 =
 \int_0^{2\pi} d^D \underline{\theta}\int_0^{2\pi} d^D \underline{\hat{\theta}}\,
 \frac{N^{D}}{(2\pi)^{D}}
  \lt[ \prod_{j=0}^{D}{p_j}^{\frac{1}{2}} e^{i(\h\theta_j-\theta_j)N_j}\rt]
 \left(\left[\prod_{k=0}^{D}\lt(\frac{p_k}{a_k^2}\right)^{p_k}\right]
 \sum_{j=0}^{D}a_j^2 e^{i(\theta_j-\hat{\theta}_j)} \right)^N.
\end{eqnarray}
This can be written as the integration of a multi-dimensional delta-function. Indeed, if the sum under the $N$-power on the r.h.s.~has modulus greater than or equal to one for $\underline{\theta}=\hat{\underline{\theta}}$ and less than one for $\underline{\theta}\neq \hat{\underline{\theta}}$, then in the large-$N$ limit the integrand will be supported on $\underline{\theta}=\hat{\underline{\theta}}$, hence will be a multi-dimensional delta function. This condition is only met if we set $a_j=\sqrt{p}_j$ for all $j=0,\ldots,D$; then for $\underline{\theta}=\hat{\underline{\theta}}$ the terms in the sum add up to one because $\sum_{j=0}^{D} p_j=1$, and otherwise they add up to a number whose modulus is less than one. This implies that
\begin{equation}
  (2\pi N)^{D}
   \lt[ \prod_{j=0}^{D}{p_j}^{\frac{1}{2}} e^{i(\h\theta_j-\theta_j)N_j}\rt]
   \left(
 \sum_{j=0}^{D}p_j e^{i(\theta_j-\hat{\theta}_j)} \right)^N\stackrel{N\to\infty}=\prod_{j=0}^{D-1}2\pi \delta(\theta_j-{\hat{\theta}}_j).
\end{equation}

Thus, the vectors $|\Psi_{\cvec{v}(\underline{\theta},\underline{a})}\rangle$ are of the same type as their finite $N$ counterparts with
\begin{equation}\label{ss}
  |\Psi_{\cvec{v}(\underline{\theta},\underline{a})}\rangle = \lim_{N\to\infty} (\sqrt{2\pi N})^D \prod_{j=0}^{D}\sqrt{a_j} e^{-i\theta_jN_j}  |{\psi}_{\cvec{v}(\underline{{{\theta}}},\underline{a})}\rangle,
\end{equation}
that is, $R_{\cvec{v}(\underline{\theta},\underline{a})}(N) = (2\pi N)^s \prod_{j=0}^{D}\sqrt{a_j} e^{-i\theta_jN_j} $. From (\ref{main}), this implies (\ref{main2}) and (\ref{ap}).

\subsection{Finite linear combinations of elementary vectors}\label{ssectflinpopov}

We consider vectors
\beq
	|\psi_N\ket = \sum_{\underline{N}\in P_N} c_{\underline{N}}
	|\Psi(\underline{N})\ket
\eeq
where each element in the set $Q_N$ of multiplets $\underline{N} = (N_0,N_1,\ldots,N_D)$ is associated to a $N$-independent multiplet $\underline{p}=(p_0,p_1,\ldots,p_j)\in Q$ with $\sum_j p_j=1$, such that $\lim_{N\to\infty} N_j/N = p_j$. We assume that $\lim_{N\to\infty} c_{\underline{N}} = c_{\underline{p}}$ exists.

The limit $N\to\infty$ of the state $\psi_N$ is relatively easy to obtain. We only have to notice that if $\underline{p}$ and $\underline{p}'$ are different multiplets (i.e.~they differ in at least one element), then, with obvious notation, for every finitely-supported operator $\Or$,
\beq
	\lim_{N\to\infty} \bra\Psi(\underline{N})|\Or|\Psi(\underline{N}')\ket = 0.
\eeq
Indeed, $\Or$ changes the number of sites with vector $v_j$ (see (\ref{psgs})) only by a finite amount, for every $j$. Since for different $\underline{p}$ and $\underline{p}'$, there are $j$'s for which $N_j$ and $N_j'$ differ by arbitrary large amounts as $N\to\infty$, the result follows. Along with (\ref{limit}), this yields
\beq\label{popovlin}
	\lim_{N\to\infty} \psi_N = \sum_{\underline{p}\in Q} |c_{\underline{p}}|^2
	\Psi(\underline{p}).
\eeq

\sect{Asymptotic behaviour of the entanglement entropy: submanifold cases}\label{sect6}

We now analyze Formula (\ref{Sn}) and Conjecture \ref{th3}, first providing verifications for various cases where the state $\psi$
corresponds to an integration on a submanifold of $\CP^D$, then generalizing the analysis to a saddle point argument for general submanifolds.

\subsection{Finite linear combinations}

For finite linear combinations, we have found the measure (\ref{mufinite}) supported on the finite set $V$ with weight $|c_{\cvec v}|^2$. In such states, the R\'enyi entropy of a block of spins of length $m$ is, according to (\ref{Sn}),
\beq\label{rgsa}
    S_n= \frc1{1-n} \log\lt(\sum_{\cvec{v}_\alpha\in V\atop\alpha=1,\ldots,n}
        \lt[\prod_{\alpha=1}^n |c_{\cvec{v}_\alpha}|^2 \rt]
        \lt[\prod_{\alpha=1}^n \cvec{v}^{\, \dagger}_\alpha\cdot\cvec{v}_{\alpha+1}\rt]^{m}\rt)
\eeq
(recall that the vectors $ \cvec{v}_\alpha$ have length one). Like for the spin $\frac{1}{2}$ case \cite{fractal}, we find, for all spins, saturation at $m$ large
\beq\label{saturation}
    \lim_{m\to\infty} S_n =
    \frc1{1-n}\log\lt(\sum_{\cvec{v}\in V} |c_{\cvec{v}}|^{2n}\rt).
\eeq
That is, the entanglement entropy of the thermodynamic limit of a finite linear combination of
 basic zero entropy states $|\psi_{\cvec{v}}\rangle$ reaches a finite maximum as
 $m \rightarrow \infty$. This maximum is obtained from the sum in (\ref{rgsa}) by taking the terms where all vectors $v_\alpha$ are aligned.

\subsection{Elementary states}

We now present a verification of Conjecture \ref{th3} in the case of the permutation symmetric states (\ref{popovstates}), using the result expressed around (\ref{sub}) and derived in Subsection \ref{ssectthermopopov}. In this and the next subsection, we will use the notation
 \begin{equation}
   \cvec{v}(\underline{\theta},\underline{p}):=\left(\prod_{k=0}^{D}e^{-i p_k \theta_k}\right)\left(
              \begin{array}{c}
                \sqrt{p_0} e^{i \theta_0} \\
                \sqrt{p_1} e^{i\theta_1} \\
                \vdots \\
                \sqrt{p_{D}} e^{i\theta_{D}} \\
              \end{array}
            \right) \qquad \text{with} \qquad \sum_{j=0}^{D}p_j=1.\label{con2}
 \end{equation}
This is essentially $\psi_{\cvec v (\underline{\theta},\underline{a})}$ with $a_j=\sqrt{p_j}$, and with an additional overall $\underline{\theta}$ and ${\underline{p}}$-dependent phase that we put for convenience.

Thanks to (\ref{main2}), we effectively have a (continuous) \textit{infinite} linear combination of the zero-entropy states (\ref{ss}). The fact that the linear combination is now infinite allows the entanglement entropy of a block to diverge as the size of the block increases (rather than saturate, as in the previous case). According to Theorem \ref{th2}, the entropy in the large-$N$ limit is simply
\begin{eqnarray}\label{rgsaq}
    S_n=\frc1{1-n} \log\left(  \int_0^{2\pi} \lt[\prod_{\alpha=1}^n \frc{d^D \underline{\theta}^\alpha}{(2\pi)^D}\rt]
      \lt[\prod_{\alpha=1}^n \cvec{v}^{\, \dagger}(\underline{\theta}^\alpha,\underline{p})\cdot\cvec{v}(\underline{\theta}^{\alpha+1},\underline{p})\rt]^{m}\right),
    \end{eqnarray}
    where we have taken the measure $\mu$ given in (\ref{sub}) (and the additional phase in (\ref{con2}) cancels out).
    Note that, in the scalar product above, the variables $\underline{p}$ are fixed and identical for every copy $\alpha$. Thus, the particular infinite linear combination of zero-entropy states considered in this example only involves the subset of states which is parametrized by the same values of $\underline{p}$, that is states with fixed magnetization. For such states the scalar product of single site vectors is given by
    \begin{equation}
\cvec{v}^{\, \dagger}(\underline{\theta}^\alpha,\underline{p})\cdot\cvec{v}(\underline{\theta}^{\alpha+1},\underline{p})= \prod_{k=0}^{D} e^{-i(\theta_k^{\alpha+1}-\theta_k^{\alpha}) p_k}\sum_{j=0}^{D} p_j e^{i(\theta_j^{\alpha+1}-\theta_j^{\alpha})}.
    \end{equation}

   As previously done for the $s=1/2$ case in \cite{fractal}, we will now carry out a detailed saddle point analysis on the integral (\ref{rgsaq}), elaborating on the results presented in \cite{fractal}. First, we expand the logarithm of the scalar product above about $\theta_j^\alpha=\theta_j^{\alpha+1}$ for all $j$ and $\alpha$
    \begin{eqnarray}
        \log\left(\ \prod_{k=0}^{D} e^{-i(\theta_k^{\alpha+1}-\theta_k^{\alpha}) p_k}\sum_{j=0}^{D} p_j e^{i(\theta_j^{\alpha+1}-\theta_j^{\alpha})}\right)&=&\frac{1}{2} \sum_{j\neq k} p_k p_j (\theta_k^{\alpha+1}-\theta_k^{\alpha})(\theta_j^{\alpha+1}-\theta_j^{\alpha})\nonumber\\
        &+& \frac{1}{2}\sum_{k=0}^{D} p_k(p_k-1) (\theta_k^{\alpha+1}-\theta_k^{\alpha})^2,
    \end{eqnarray}
neglecting terms of order 3 and above. Note that this is purely real and has no first-order term; this is thanks to our choice of phase factor in (\ref{con2}). This
    yields
   \begin{equation}
   \prod_{\alpha=1}^n \lt[\cvec{v}^{\, \dagger}(\underline{\theta}^\alpha,\underline{p})\cdot\cvec{v}(\underline{\theta}^{\alpha+1},\underline{p})\rt]^{m} =
     e^{\sum\limits_{\alpha=1}^n \left[\frac{m}{2} \sum\limits_{j\neq k}p_k p_j (\theta_k^{\alpha+1}-\theta_k^{\alpha})(\theta_j^{\alpha+1}-\theta_j^{\alpha})+ \frac{m}{2}\sum\limits_{k=0}^{D} p_k(p_k-1) (\theta_k^{\alpha+1}-\theta_k^{\alpha})^2\right] }.\label{res}
   \end{equation}
As we can see from (\ref{rgsaq}) and (\ref{res}) although the scalar products depend on the variables $\theta_{D}^j$ these variables are not integrated on in (\ref{rgsaq}) (recall the definition (\ref{dtheta})). They represent the fact that the vectors (\ref{con2}) are fixed up to a phase. We choose once more $\theta_{D}^\alpha=0$.

Let us plug (\ref{res}) into (\ref{rgsaq}) and change variables to $t_{j}^\alpha=\sqrt{m} p_j (\theta_j^{\alpha}-\theta_j^1)$ for $\alpha=2,\ldots,n$ and $t_j= \theta_j^1$ for $j=0,\ldots,D-1$. With this change of variables the integrand does not depend on the variables $t_j$ so that $D$ of the integrals can be computed, contributing a factor $(2\pi)^{D}$ which is canceled by the same factor in the denominator. The remaining integrals are of Gaussian type and can be computed using standard techniques:
\begin{eqnarray}\label{rgsa22}
    S_n&=&\frc1{1-n} \log\left[{\left(m^s \prod_{j=0}^{D-1} p_j\right)^{1-n}}\prod_{\alpha=2}^n \frac{1}{(2\pi)^{D}}\int_{-\infty}^{\infty} d^D \underline{t}^\alpha \, e^{
       \frac{1}{2} \sum\limits_{j\neq k} (t_k^2 t_j^2+t_k^n t_j^n)+ \frac{1}{2}\sum\limits_{k=0}^{D-1} \frac{p_k-1}{p_k} ((t_k^2)^2+(t_k^n)^2) } \right.\nonumber\\
    &&  \qquad\qquad\left. \times e^{\sum\limits_{\alpha=2}^{n-1}\left[\frac{1}{2} \sum\limits_{j\neq k}(t_k^{\alpha+1}-t_k^{\alpha})(t_j^{\alpha+1}-t_j^{\alpha})+ \frac{1}{2}\sum\limits_{k=0}^{D-1} \frac{p_k-1}{p_k} (t_k^{\alpha+1}-t_k^\alpha)^2 \right]}\right].
    \end{eqnarray}
    Therefore,
    \begin{equation}\label{dom}
        S_n\sim s\log(4\pi^2 m) +  \log \left(\prod_{j=0}^{D-1} p_j\right) + \frac{\log F(n)}{1-n},
    \end{equation}
    with
    \begin{equation}
      F(n) = \prod_{\alpha=2}^n\int_{-\infty}^{\infty} d^D \underline{t}^\alpha \, e^{\sum\limits_{\alpha=2}^{n}\left[
       \sum\limits_{j\neq k} t_k^\alpha t_j^\alpha+ \sum\limits_{k=0}^{D-1} \frac{p_k-1}{p_k}
       (t_k^\alpha)^2\right]-\sum\limits_{\alpha=2}^{n-1}\left[\frac{1}{2} \sum\limits_{j\neq k}(t_k^{\alpha+1}t_j^{\alpha}+ t_k^{\alpha}t_j^{\alpha+1}) + \sum\limits_{k=0}^{D-1} \frac{p_k-1}{p_k}
       t_k^\alpha t_k^{\alpha+1})\right]}.
    \end{equation}

    This expression clearly shows the leading $s\log m$ behaviour of the entropy for $m$ large already given in \cite{pop1}. The additional terms can be obtained by computing the Gaussian integrals using the general result
\begin{equation}
\int_{-{\infty}}^{\infty} e^{-\frac{1}{2}\sum_{i,j=1}^M A_{ij}x_i
x_j + \sum_{i=1}^M B_i x_i}d^M x= \sqrt{\frac{(2\pi)^M}{\det
A}}e^{\frac{1}{2}\cvec{B}^{t}\cdot A^{-1} \cdot \cvec{B}},
\end{equation}
where, in our case, the number of variables is $M=D(n-1)$ and
$\cvec{B}=0$. The $\det A$ can be computed in general to:
\begin{equation}
    \det A= n^{D} \left(-\sum_{k=0}^{D-2}(D-1-k)\sigma_k + \sigma_{D}\right)^{n-1},
\end{equation}
where $\sigma_k$ are elementary symmetric polynomials on the variables $\{\frac{1-p_0}{p_0},\ldots, \frac{1-p_{D-1}}{p_{D-1}}\}$. Due to  the property $\sum_{i=0}^{D}p_i=1$ it turns out that the expression above is equivalent to,
\begin{equation}
    \det A= n^{D} \left(\frac{p_{D}}{\prod_{j=0}^{D-1}p_j} \right)^{n-1},
\end{equation}
and therefore,
\begin{equation}
    F(n)=\frac{1}{(2\pi)^{s(1-n)} n^s}\left(\frac{\prod_{j=0}^{D-1}p_j}{p_{D}}\right)^{\frac{n-1}{2}}.
\end{equation}
This immediately gives (\ref{renyi2}) and (\ref{vn3}). The expression (\ref{vn3}) exactly agrees with the result obtained in \cite{pop2}. These results may also be compared to the expressions found for the $s=\frac{1}{2}$ case and the particular state of the Ferromagnetic XXX chain considered in \cite{permutation}. We referred to this state as being associated to a ``great circle" on the two-dimensional sphere (Bloch sphere). In the present language, the state in question is the zero-magnetization permutation symmetric state corresponding to setting $p_0=p_1=\frac{1}{2}$. For these particular values we recover the results given in equation (5.73) of \cite{permutation}, as expected.

\subsection{Integration on $\CP^D$}

The states considered in the previous subsection exhibit logarithmic scaling of both the R\'enyi and von Neumann entanglement entropy as $s\log m$ for large blocks of size $m$. Reviewing the methodology employed there, it is straightforward to generalize this result to other linear combinations of the basic states (\ref{ss}), with integrations over $\underline{\theta}$ and summation over $\underline{p}$ (recall that $a_j^2=p_j$). A simple example of such states are the states (\ref{popovlin}) of Subsection \ref{ssectflinpopov}, obtained as the thermodynamic limit of a finite linear combination of states (\ref{psgs}). Another simple example is obtained if we choose a number $r$ of the parameters $\theta_j$ to be fixed. This would reduce the number of integrals involved in (\ref{rgsa}) from $D n$ to $(D-r) n$. Thus, carrying out a similar change of variables as in the previous section for the remaining variables the entropy would now diverge as $(s-\frac{r}{2})\log m$. We did not obtained these states as thermodynamic limits of sequences of finite-$N$ states, but the second part of Theorem \ref{th1} guarantees that they do occur.  Given that $r=0,\ldots,D$ this provides a large number of different scaling behaviours which the entropy will exhibit, depending on the choice of the initial state. These cases correspond to the measure $\mu$ in (\ref{psiO}) being supported on various submanifolds, and not only part I of Conjecture \ref{th3} hold for these cases, but also Part II, although this requires some more analysis (see the general analysis in the next subsection).
In addition, if we further give different weights to the various basic states $|\psi_{\cvec{v}(\underline{\theta},\underline{p})}\rangle$ in the linear combination, we will find that this does not change the leading behaviour of the entropy for large $m$ but will change next-to-leading (constant) corrections and $n$-dependent corrections, in the case of the R\'enyi entropy.

More generally, it is also possible to construct linear combinations of the states $|\psi_{\cvec{v}(\underline{\theta},\underline{p})}\rangle$  whose entropy for large $m$ is larger than in any of the cases considered in the previous paragraph. This may be achieved by allowing, in the thermodynamic limit, the parameters $p_j$, which where fixed in the previous section, to vary in a continuous fashion (i.e.~by taking infinite linear combinations, or integrals, over the $p_j$ as well). We have not shown that these occur as some thermodynamic limits, but Theorem \ref{th1} guarantees it. If all $D=2s$ independent parameters are allowed to vary one can show that the entropy will then scale as $2s\log m$. By a similar argument as that above, if $r$ parameters $p_i$ are fixed the large-block entropy will then be dominated by a term $(D-\frac{r}{2})\log m$. We can therefore see that playing with the number of parameters $\underline{\theta}$ and $\underline{p}$ which are allowed to vary, the entropy scales as
$\frac{d}{2}\log m$ with $d=0,\ldots,4s$.

We now generalize the analysis of the previous section to the situation when both sets  $\underline{\theta}$ and $\underline{p}$ are integrated over. In addition, we will consider the most general situation, when we have an arbitrary linear combination of states, parametrized by functions $g(\underline{\theta}^\alpha,\underline{\phi}^\alpha)\geq 0$:
 \begin{eqnarray}\label{rgsa2}
    S_n=\frc1{1-n} \log\left(\prod_{\alpha=1}^n \int_0^{2\pi} d^D \underline{\theta}^\alpha
    \int_{S^{D}} d^D \underline{\phi}^\alpha
      \prod_{\alpha=1}^n g(\underline{\theta}^\alpha,\underline{\phi}^\alpha) \lt[\cvec{v}^{\, \dagger}(\underline{\theta}^\alpha,\underline{\phi}^\alpha)\cdot\cvec{v}(\underline{\theta}^{\alpha+1},\underline{\phi}^{\alpha+1})\rt]^{m}
      \right),
    \end{eqnarray}
 where $S^{D}$ is the $D$-dimensional sphere. Here, instead of using our original variables $\underline{p}^\alpha$ we have chosen  to use a
 natural parametrization of the sphere in terms of angles $\underline{{\phi}}^\alpha$ as given in (\ref{spherical}) with the restriction (\ref{intervals}) and up to the usual identification $p_j^\alpha=(a_j^\alpha)^2$.

 Comparing (\ref{rgsa2}) to the general formula (\ref{Sn}), we have that the measure $\mu$ has been chosen so that
 \begin{equation}\label{mumu}
    d\mu(\cvec{v}(\underline{\theta},\underline{\phi}))= g(\underline{\theta},\underline{\phi}) \,d^D \underline{\theta} d^D \underline{\phi}.
 \end{equation}

  Our starting point as before is the scalar product of single normalized vectors
    \begin{equation}
\cvec{v}^{\, \dagger}(\underline{\theta}^\alpha,\underline{\phi}^\alpha)\cdot\cvec{v}(\underline{\theta}^{\alpha+1},\underline{\phi}^{\alpha+1})= \prod_{k=0}^{D} e^{-i(\theta_k^{\alpha+1}p_k^{\alpha+1}-\theta_k^{\alpha}p_k^\alpha)}\sum_{j=0}^{D} \sqrt{p_j^{\alpha+1}  p_j^{\alpha}}e^{i(\theta_j^{\alpha+1}-\theta_j^{\alpha})}. \label{product}
    \end{equation}
Expanding the logarithm of this scalar product about $\theta_j^\alpha=\theta_j^{\alpha+1}$ and $\phi_j^\alpha=\phi_j^{\alpha+1}$ for all $j$ and $\alpha$ we find, up to second order,
    \begin{eqnarray}
        && \log\left(\cvec{v}^{\, \dagger}(\underline{\theta}^\alpha,\underline{\phi}^\alpha)\cdot\cvec{v}(\underline{\theta}^{\alpha+1},\underline{\phi}^{\alpha+1})\right)=\frac{1}{2} \sum_{j\neq k} p_k^1 p_j^1 (\theta_k^{\alpha+1}-\theta_k^{\alpha})(\theta_j^{\alpha+1}-\theta_j^{\alpha})\nonumber\\
        &+& \frac{1}{2}\sum_{k=0}^{D} p_k^1(p_k^1-1) (\theta_k^{\alpha+1}-\theta_k^{\alpha})^2 - \frac{1}{2}\sum_{k=0}^{D-1} \left(\prod\limits_{p=0}^{k-1} \sin^2 \phi_p^1 \right)(\phi_k^{\alpha+1}-\phi_k^\alpha)^2
            - i\omega^\alpha
    \end{eqnarray}
    where we used
 \beq
 	\sum_{j=0}^D \frc{(p_j^{\alpha+1}-p_j^\alpha)^2}{4p_j^\alpha}
	= \sum_{k=0}^{D-1} \left(\prod\limits_{p=0}^{k-1} \sin^2 \phi_p^1 \right)(\phi_k^{\alpha+1}-\phi_k^\alpha)^2,
 \eeq
valid to second order. In this expression we have used that to second order, the coefficients of the quadratic terms above can be written in terms of the variables $p_j^1$ only (or equivalently $\phi_j^1$), for each value of $j$. The real quantity $\omega^\alpha$ is given by
\beq\label{omega}
	\omega^\alpha = \sum_{j=0}^D \lt[(p^{\alpha+1}_j-p^\alpha_j)\theta_j^\alpha +\frc12 (p_j^{\alpha+1}-p_j^\alpha)(\theta^{\alpha+1}_j-\theta^\alpha_j)\rt].
\eeq

Contrary to the case of the previous subsection, the pure imaginary part $-i\omega^\alpha$ cannot in general be canceled by a choice of phases of the vectors $\cvec{v}(\underline{\theta}^\alpha,\underline{\phi}^\alpha)$. In particular, the quantity
\beq
	\Omega = \sum_{\alpha=1}^n \omega^\alpha
\eeq
is not zero for any $n\geq 3$ (it is zero for $n=2$). However, since the first term under the summation symbol in (\ref{omega}), when summed over $\alpha$, is invariant under $\theta_j^\alpha \mapsto \theta_j^\alpha + q_j$ for any $q_j$, the quantity $\Omega$ is in fact a function of $\theta_j^{\alpha+1}-\theta_j^{\alpha}$ and $p^{\alpha+1}_j-p^\alpha_j$ (for $j=0,\ldots,D$, $\alpha=1,\ldots,n$) only. From (\ref{omega}) it is clear that it is quadratic in these differences.

Putting everything together we have
   \begin{eqnarray}
   \prod_{\alpha=1}^n \lt[\cvec{v}^{\, \dagger}(\underline{\theta}^\alpha,\underline{\phi}^\alpha)\cdot\cvec{v}(\underline{\theta}^{\alpha+1},\underline{\phi}^{\alpha+1})\rt]^{m} &=&
     e^{\sum\limits_{\alpha=1}^n \left[\frac{m}{2} \sum\limits_{j\neq k}p_k^1 p_j^1 (\theta_k^{\alpha+1}-\theta_k^{\alpha})(\theta_j^{\alpha+1}-\theta_j^{\alpha})+ \frac{m}{2}\sum\limits_{k=0}^{D-1} p_k^1(p_k^1-1) (\theta_k^{\alpha+1}-\theta_k^{\alpha})^2\right] }\nonumber\\
     && e^{- \frac{m}{2}\sum\limits_{\alpha=1}^n\sum\limits_{k=0}^{D-1} \left(\prod\limits_{p=0}^{k-1} \sin^2 \phi_p^1 \right)(\phi_k^{\alpha+1}-\phi_k^\alpha)^2 } e^{-mi\Omega}.\label{res2}
   \end{eqnarray}
We may now substitute this expression into (\ref{rgsa2}). By symmetry under exchange of copies of the integration measure in (\ref{rgsa2}), we find that the result of the integral in (\ref{rgsa2}) is real. Hence, we may replace, in the integrand, the factor $e^{-mi\Omega}$ by its real part $\cos(m\Omega)\approx e^{-m^2\Omega^2/2}$ (in the saddle point approximation). Since $\Omega$ is quadratic in the differences $\theta_j^{\alpha+1}-\theta_j^{\alpha}$ and $p^{\alpha+1}_j-p^\alpha_j$, then $\Omega^2$ is quartic. This shows that the phase factor $e^{-mi\Omega}$ in (\ref{res2}) can be neglected in the saddle point approximation of the integral.

We now proceed as in the previous section. The integrals in the variables $\theta_j^\alpha$ may in fact be carried out exactly as before, with the only difference that the parameters $p_j^1$ are not constants in the present case and therefore their product must be kept inside the integral. After the change of variables
$t_j^{\alpha}=\sqrt{m}p_j^1 (\theta_j^{\alpha}-\theta_j^1)$ and $h_j^\alpha=\sqrt{m}(\phi_j^\alpha-\phi_j^1)$ for $\alpha\geq 2$ and $t_j^1=\theta_j^1$, $h_j^1=\phi_j^1$ we have:
\begin{eqnarray}\label{rgsa3}
    S_n &=&2s\log \frac{m}{\sqrt{2 \pi}}  -\frac{s \log n}{1-n}+  \frc1{1-n} \log\left(
    \int_0^{2\pi} d \underline{{t}}^1 \int_{S^{D}} d \underline{h}^1  \left({\prod_{j=0}^{D} p_j^1}\right)^{\frac{1-n}{2}}g(\underline{t}^1,\underline{h}^1)^n \right.
     \\
     & & \left.\int_{-\infty}^\infty  \prod_{\alpha=2}^n d \underline{h}^\alpha e^{- \frac{1}{2}\sum\limits_{\alpha=2}^{n-1}\sum\limits_{k=0}^{D-1} \left(\prod\limits_{p=0}^{k-1} \sin^2 h_p^1 \right)(h_k^{\alpha+1}-h_k^\alpha)^2-\frac{1}{2}\sum\limits_{k=0}^{D-1} \left(\prod\limits_{p=0}^{k-1} \sin^2 h_p^1 \right)((h_k^{2})^2+(h_k^n)^2)}\right).\nonumber
    \end{eqnarray}
The integrals in $h_j^\alpha$ for $\alpha\geq 2$ can be easily carried out as they are of the standard Gaussian type. Moreover, there is factorization for each fixed $j$. Integrating we obtain:
\begin{equation}
   \frac{(2\pi)^{s(n-1)}}{n^s \left(\prod\limits_{k=0}^{D-1}\prod\limits_{p=0}^{k-1} \sin h_p^1\right)^{n-1} }.
\end{equation}
substituting into (\ref{rgsa3}) after expressing the product $\prod_{j=0}^{D} p_j^1$ in terms of the angles $\underline{h}_j^1$ we find
\begin{eqnarray}\label{rgsa33}
    S_n &=&2s\log \frac{m}{2\pi}  -\frac{2 s \log n}{1-n}
     \nonumber \\
     & +&  \frc1{1-n} \log\left(
    \int_0^{2\pi} d \underline{{t}}^1\int_{S^{D}} d \underline{h}^1 g(\underline{t}^1,\underline{h}^1)^n \left(\prod_{k=0}^{D-1}{\frac{1}{2}\sin(2 h_k^1)}{\prod_{p=0}^{k-1} \sin^2 h_p^1}\right)^{1-n}\right),
    \end{eqnarray}
    Hence, the von Neumann entanglement entropy takes the form
    \begin{eqnarray}\label{vn}
    S_1 &=&2s\log\frac{ e m}{2\pi}
   -
   \int_0^{2\pi} d \underline{{t}}^1\int_{S^{D}} d \underline{h}^1 g(\underline{t}^1,\underline{h}^1) \log \left(\frac{g(\underline{t}^1,\underline{h}^1)}{\prod_{k=0}^{D-1}{\frac{1}{2}\sin(2 h_k^1)}{\prod_{p=0}^{k-1} \sin^2 h_p^1}}\right),
    \end{eqnarray}
    where we have used the fact that
    \begin{equation}
        {\int_0^{2\pi} d \underline{{t}}^1\int_{S^{D}} d \underline{h}^1 g(\underline{t}^1,\underline{h}^1)}=1.
    \end{equation}
   In order to compare this result to the general formula (\ref{Snass3}) we need to find the relationship between our function $g(\underline{t}^1,\underline{h}^1)$ and the function defined in Conjecture \ref{th3}.II. Recall that the defining property for this function is
   \begin{equation}
    d\mu(\cvec{v}(\underline{t},\underline{h}))=d^{2D} \cvec{v} f(\underline{t},\underline{h}),
   \end{equation}
   and $d^{2D} \cvec{v}$ is defined in (\ref{volelem}). Therefore,
   \begin{equation}\label{g}
   f(\underline{t}^1,\underline{h}^1)=\frac{g(\underline{t}^1,\underline{h}^1)}{2^{D}\prod\limits_{k=0}^{D-1}{\sin(2 h_k^1)}{\prod\limits_{p=0}^{k-1} \sin^2 h_p^1}},
   \end{equation}
   and requiring that
   \begin{equation}
    \int d^{2D} \cvec{v}  f(\underline{t}^1,\underline{h}^1)= \int_0^{2\pi} d \underline{{t}}^1\int_{S^{D}} d \underline{h}^1 g(\underline{t}^1,\underline{h}^1)=1,
   \end{equation}
   the von Neumann entropy becomes
    \begin{eqnarray}\label{vn2}
    S_1 =2s\log\frac{e m}{8\pi}
   -
  \int d^{2D}\cvec{v}\,f(\underline{t}^1,\underline{h}_1) \log f(\underline{t}^1,\underline{h}^1),
    \end{eqnarray}
as expected. Similarly, the R\'enyi entropy is that obtained from (\ref{Snass2}) with $d=4s$.

\subsection{General case}

The argument in the general submanifold case is a consequence of a similar saddle-point approximation as that done above. Let $x_j$ be real coordinates on the submanifold (formally representing all coordinate patches), with $j=1,\ldots,d$ where $d$ is the dimension of the manifold. Then the state that we are considering is
\beq
	\psi = \int d^d\underline{x}\,\sqrt{\eta(\underline{x})} \,f(\underline{x})\,\psi_{\cvec{v}(\underline{x})}
\eeq
where $\eta(\underline{x})$ is the determinant of the metric, at the point $\underline{x}$ on the submanifold, induced by the Study-Fubini metric on $\CP^D$; and where $\cvec{v}(\underline{x})$ is the homogeneous coordinate on $\CP^D$ corresponding to the point $\underline{x}$. Hence according to (\ref{Sn}) we wish to evaluate, in the large-$m$ limit, the quantity
 \begin{eqnarray}\label{rgsa6}
    S_n=\frc1{1-n} \log\lt[\int\left(\prod_{\alpha=1}^n
     d^d\underline{x}^\alpha\,\sqrt{\eta(\underline{x}^\alpha)} \,f(\underline{x}^\alpha)\rt)\lt(
      \prod_{\alpha=1}^n \cvec{v}^{\, \dagger}(\underline{x}^\alpha)\cdot\cvec{v}(\underline{x}^{\alpha+1})
      \right)^m\rt].
    \end{eqnarray}

Note first that (\ref{res2}) implies, to quadratic order (with assumed normalization $\cvec{v}^{\, \dag}_\alpha\cdot \cvec{v}_\alpha = 1$),
\beq
\prod_{\alpha=1}^n \cvec{v}^{\, \dagger}_\alpha \cdot \cvec{v}_{\alpha+1} = \prod_{\alpha=1}^n \lt|\cvec{v}^{\, \dagger}_\alpha \cdot \cvec{v}_{\alpha+1} \rt|
e^{-i\Omega}
\eeq
where $\Omega$ is quadratic in the difference of the coordinates of $\cvec{v}_\alpha$ and $\cvec{v}_{\alpha+1}$ on $\CP^D$ (these are coordinate-independent statements).
Further, the fact that the integration measure in (\ref{Sn}) is symmetric under exchange of copies implies, as in the argument after (\ref{res2}), that $e^{mi\Omega}$ can be neglected under the integral in (\ref{Sn}) in the saddle point approximation. Hence we may replace in (\ref{Sn})
\beq\label{replacement}
	\prod_{\alpha=1}^n \cvec{v}^{\, \dagger}_\alpha \cdot \cvec{v}_{\alpha+1} \mapsto e^{-\frc18 \sum_{\alpha=1}^n D(\cvec{v}_\alpha,\cvec{v}_{\alpha+1})^2}
\eeq
where we use the distance on $\CP^D$ given by (\ref{distCP}). The integration in (\ref{rgsa6}), in the large-$m$ limit, is then performed along entirely similar lines as those of the previous subsections. We use
\[
	D(\cvec{v}\lt(\underline{x}^\alpha),\cvec{v}(\underline{x}^{\alpha+1})\rt)^2 = \sum_{j,k} \eta_{j,k}(\underline{x}^1) (x_j^{\alpha+1}-x_j^{\alpha})
(x_k^{\alpha+1}-x_k^{\alpha})
\]
(to quadratic order), where $\eta_{j,k}(\underline{x}^1)$ is the metric on the submanifold, here evaluated at the coordinates $\underline{x}^1$ belonging to the copy $\alpha=1$. After the change of variable $t_j^\alpha = \sqrt{m} (x_j^\alpha-x_j^1)$, $\alpha\geq 2$ and $t_j^1 = x_j^1$, the result of the Gaussian integration over $t_j^\alpha$, $\alpha\geq 2$ is (\ref{Snass2}).

\sect{Asymptotic behaviour of the entanglement entropy: fractal subsets}\label{sect7}

In this section we complete the verification of Part I of Conjecture \ref{th3} by providing an explicit example of a case where $\psi$ corresponds to a measure over a fractal subset of $\CP^D$ with $D=1$ (i.e.~spin $s=1/2$). We then give the general argument for integrations over fractals that was briefly presented (in the special case $s=1/2$) in \cite{fractal}.

\subsection{A Cantor set on the Bloch Sphere}

Let us consider the entanglement entropy for $s=\frac{1}{2}$ and provide an explicit example in which the geometric dimension $d$ is a
fractal dimension. In our example this will be the fractal dimension of the well-known Cantor set (see e.g.~\cite{chaos}).
 Here we identify the Boch sphere $S^2$ with $\CP^1$ in the usual way, via
\[
	\cvec v = \lt(\begin{matrix} \sqrt{1+z} \\ \sqrt{1-z}\, e^{i\theta}\end{matrix}\rt),\quad x+iy = \sqrt{1-z^2}\,e^{i\theta}
\]
where $(x,y,z)\in S^2\subset \R^3$, and $\cvec v$ is a homogeneous coordinate on $\CP^1$.

Consider once more our expression for the entropy
(\ref{Sn}). Let $|c_{\cvec v}|^2>0$ for every $\cvec v$ in a finite subset of unit vectors, and consider the associated quantum state
\begin{equation}\label{qs}
    \psi_{\{c_{\cvec{v}}\}}:=\sum_{\cvec{v}} |c_{\cvec{v}}|^2\psi_{\cvec{v}},
\end{equation}
Its entanglement entropy is
\beq\label{rgsa5}
    S_n= \frc1{1-n} \log\lt(\sum_{\{\cvec{v}_\alpha\}}
        \lt[\prod_{\alpha} |c_{\cvec{v}_\alpha}|^2 \rt]
        \lt[\cvec{v}_\alpha^\dagger \cdot \cvec{v}_{\alpha+1}\rt]^{m}
        \rt),
\eeq
where, as in the previous subsection, we used the normalization $|\cvec v_\alpha|^2=1$. Note that the argument leading to (\ref{replacement}) holds true for any measure of integration. Hence, in general to quadratic order, we may make the replacement (here specialized to $s=1/2$)
\beq\label{asympoverlap}
    \prod_{\alpha=1}^n \cvec{v}_\alpha^\dagger \cdot \cvec{v}_{\alpha+1}
    \mapsto \exp\lt[-\frac{1}{8}\sum_{\alpha=1}^n |\vec{v}_{\alpha+1}
    -\vec{v}_\alpha|^2\rt]
\eeq
where $\vec{v}_\alpha\in S^2$ is the coordinate on $S^2$ representing the homogeneous coordinate $\cvec v_\alpha\in \CP^1$.

Consider instead a state in a linear combination supported on the Cantor set (the method considered here can be applied to other similar fractal sets). For definiteness, we consider a Cantor set on half of a great circle on the Bloch sphere, as shown in Fig.~1.
 \begin{figure}[h!]
\begin{center}
\includegraphics[angle=0,height=6cm, width=6cm]{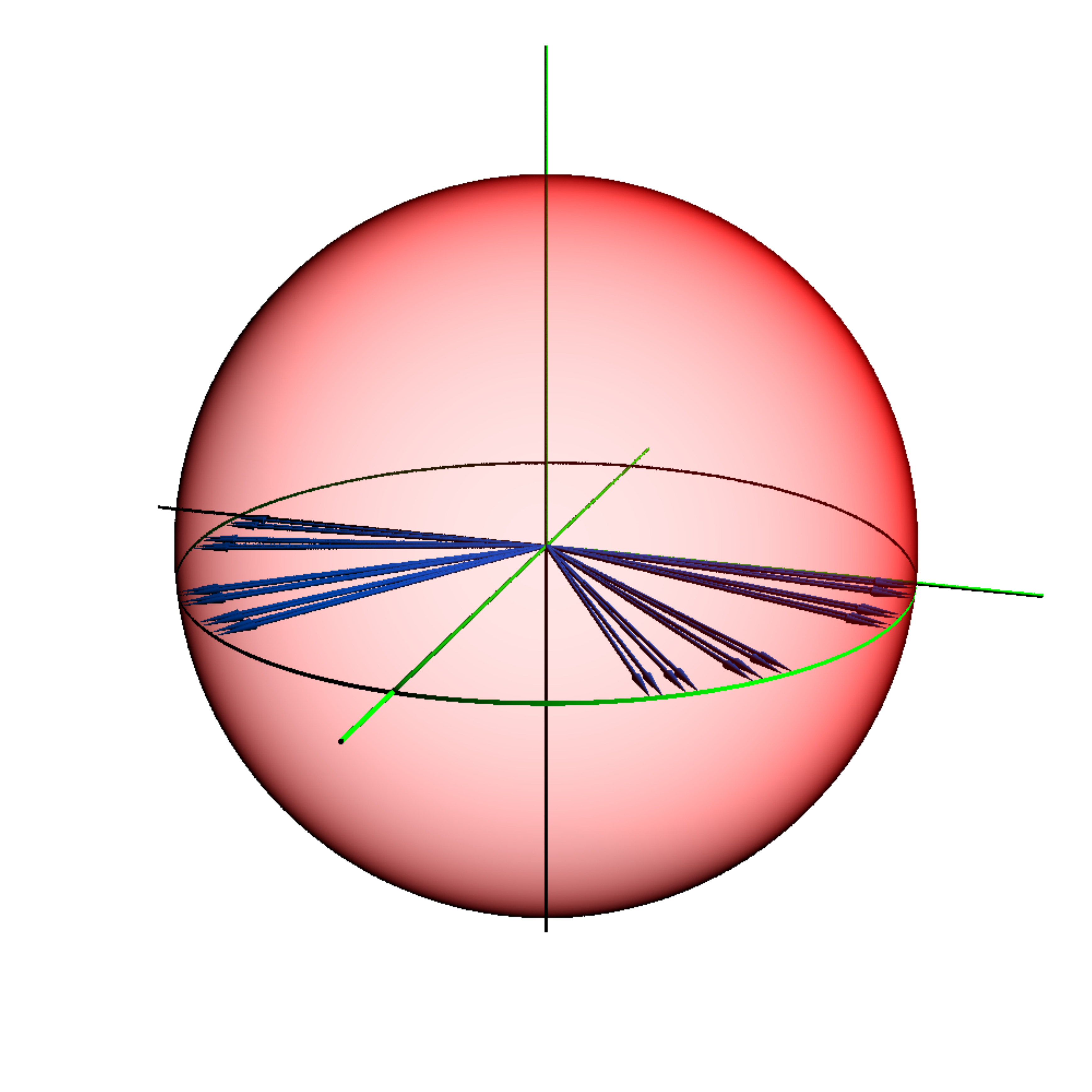}
\caption{A Cantor set (pointed by the blue arrows) on half of the equator (green circle) of the Bloch sphere (semi-transparent red). The points are computed by employing the usual iterative definition of the Cantor set up to 5 iterations, and the
map (\ref{map}).}
\end{center}
\end{figure}
 The great circle that we consider is that with $z=0$. The Cantor set $C$ on the unit interval $[0,1]$ is the limit
\[
	C = \lim_{k\to\infty} { C}_k
\]
where the finite sets ${ C}_k:k=0,1,2,\ldots$ are defined recursively by
\[
	{ C}_{k+1} = \frc13 \lt({ C}_{k} \cup (2+{ C}_k)\rt),\quad
	{ C}_0 = \{0\}
\]
(here we use the general notation $a+b\{q_1,q_2,\ldots\}=\{a+bq_1,a+bq_2,\ldots\}$).
We embed it on the Bloch sphere via the map $\phi:[0,1]\to S^2$, defined by
\beq
	\phi(x) = (\cos\pi x,\sin\pi x,0). \label{map}
\eeq
The quantum state $\psi_{\rm Cantor}$ is a Hausdorff integral over the set $\phi({ C})\subset S^2$; for simplicity, we take it with constant coefficients. We denote the Hausdorff measure on $C$ as $\mu_C$. Hence, we consider
\beq
	\psi_{\rm Cantor} = \int_{\phi( C)} d\mu_C(\cvec{v})\,\psi_{\cvec{v}}\quad
	\mbox{with measure normalized as}
	\quad
	\int_{\phi( C)} d\mu_C(\cvec{v}) = 1.
\eeq

The measure $\mu_C$ can be defined recursively following the recursive definition of the Cantor set itself. We note that in the finite set ${ C}_k$, at step $k$ of the recursive definition, there are exactly $2^k$ points. Hence the step-$k$ quantum state $\psi_k$ associated to ${ C}_k$ with constant coefficients is
\beq
	\psi_k = \frc1{2^k} \sum_{\cvec{v}\in \phi({ C}_k)} \psi_{\cvec{v}}.
\eeq
Then $\psi_{\rm Cantor}$ is defined by taking the limit:
\beq
	\psi_{\rm Cantor} = \lim_{k\to\infty} \psi_k.
\eeq
One can show that this limit exists weakly (i.e.~in the local-operator topology on infinite-$N$ (thermodynamic) states).

Consider now the entanglement entropy $S_n=S_n(\psi)$ as a function of the state $\psi$. It is clearly continuous as it is a continuous function of the evaluation of $\psi$ on a finitely-supported operator, see (\ref{here}). Hence in order to evaluate the entanglement entropy of $\psi_{\rm cantor}$, we may use formula (\ref{rgsa}) and $\lim S_n(\psi) = S_n (\lim \psi)$, giving
\beq
	S_n(\psi_{\rm Cantor}) = \lim_{k\to\infty} S_n(\psi_k)
	=
	\frc1{1-n} \log\lt(\lim_{k\to\infty}
        \frc1{2^{kn}}
	\sum_{\{\cvec{v}_\alpha\in\phi({ C}_k)\}}
        \lt[\prod_{\alpha=1}^n \bra\psi_{\cvec{v}_\alpha}|\psi_{\cvec{v}_{\alpha+1}}\ket\rt]^{m}\rt).
\eeq

At large $m$, the only terms that give important contributions are those for which all overlaps $\bra\psi_{\cvec{v}_\alpha}|\psi_{\cvec{v}_{\alpha+1}}\ket$ (for all $\alpha$) are near to 1. When such an overlap is near to 1, we can use the approximation (\ref{asympoverlap}). Further, for two vectors $\vec{v} = \phi(x)$ and $\vec{w}=\phi(y)$ near to each other on $S^2$ (i.e.~with $|x-y|$ small), we can approximate $|\vec{v}-\vec{w}|^2 \approx \pi^2 |x-y|^2$. Hence, for the purpose of evaluating the large-$m$ leading logarithmic behaviour of $S_n(\psi_{\rm Cantor})$ we can use
\beq
	\sum_{\{\cvec{v}_\alpha\in\phi({ C}_k)\}}
        \lt[\prod_{\alpha=1}^n \bra\psi_{\cvec{v}_\alpha}|\psi_{\cvec{v}_{\alpha+1}}\ket\rt]^{m} \approx
	\sum_{\{x_\alpha\in{ C}_k\}}
        \exp\lt[-\frc{\pi^2m}8 \sum_{\alpha=1}^n |x_{\alpha+1}-x_\alpha|^2\rt]
\eeq

Now we simply use the recursive definition of ${\cal C}_k$ and the fact that the large-$k$ limit exists. We have
\beqa
	Q_m&:=&
	\lim_{k\to\infty}
        \frc1{2^{kn}}
	\sum_{\{x_\alpha\in{ C}_k\}}
        \exp\lt[-\frc{\pi^2m}8 \sum_{\alpha=1}^n |x_{\alpha+1}-x_\alpha|^2\rt] \n
        &=&
	\lim_{k\to\infty}
        \frc1{2^{(k+1)n}}
	\sum_{\{x_\alpha\in{ C}_{k+1}\}}
        \exp\lt[-\frc{\pi^2m}8 \sum_{\alpha=1}^n |x_{\alpha+1}-x_\alpha|^2\rt] \n
        &=&
	\frc1{2^n} \lim_{k\to\infty}
        \frc1{2^{kn}}
	\sum_{\{x_\alpha\in\frc13 \lt({ C}_{k} \cup (2+{ C}_k)\rt)\}}
        \exp\lt[-\frc{\pi^2m}8 \sum_{\alpha=1}^n |x_{\alpha+1}-x_\alpha|^2\rt] \n
        &=&
	\frc1{2^n} \lim_{k\to\infty}
        \frc1{2^{kn}}
	\sum_{\{x_\alpha\in{ C}_{k} \cup (2+{ C}_k)\}}
        \exp\lt[-\frc{\pi^2(m/9)}8 \sum_{\alpha=1}^n |x_{\alpha+1}-x_\alpha|^2\rt] \label{eqQm1}
\eeqa
For the large-$m$ power law asymptotic, we can write in (\ref{eqQm1})
\[
	\sum_{\{x_\alpha\in{ C}_{k} \cup (2+{ C}_k)\}}
	\approx
	\sum_{\{x_\alpha\in{ C}_{k}\}} +
	\sum_{\{x_\alpha\in 2+{ C}_k\}}.
\]
Indeed, whenever a nonempty proper subset of $\{x_1,\ldots,x_n\}$ is in ${ C}_k$ and the rest is in $2+{ C}_k$, there will be finite contributions in the exponential which make it decay exponentially at large $m$. Since both sums then give the same contribution, we obtain
\beqa
	Q_m&\approx &
	\frc1{2^{n-1}} \lim_{k\to\infty}
        \frc1{2^{kn}}
	\sum_{\{x_\alpha\in{\cal C}_{k}\}}
        \exp\lt[-\frc{\pi^2(m/9)}8 \sum_{\alpha=1}^n |x_{\alpha+1}-x_\alpha|^2\rt] \n
        &=& \frc1{2^{n-1}} Q_{m/9}
\eeqa
Defining $q_\ell = \log_{2^{n-1}} Q_{9^\ell}$ (for $\ell$ large) we then find the recursion relation
\beq
	q_{\ell+1} = q_\ell - 1.
\eeq
Since this must be true for a continuum of $\ell$ (because we are looking at the asymptotic region where the integer $9^\ell$ is large) and since $q_\ell$ is monotonously decreasing with $\ell$, the unique solution is
\beq
	q_\ell = r - \ell
\eeq
for some number $r$. Hence,
\beq
	Q_m = R\, m^{-\frc{n-1}2 \frc{\log 2}{\log 3}}
\eeq
(for $m$ large) for some other number $R$.

From this, we obtain at large $m$
\beq
	S_n(\psi_{\rm Cantor}) \approx \frc1{1-n} \log \lt(m^{-\frc{n-1}2 \frc{\log 2}{\log 3}}\rt) =
	\frc12 \frc{\log 2}{\log 3}\log m = \frc{d_{\rm Cantor}}2 \log m
\eeq
where $d_{\rm Cantor} = \log2/\log3$ is the fractal dimension of the Cantor set. This agrees with (\ref{Snass1}).

\subsection{General argument}

As is clear form the calculation above, a crucial feature of
the Hausdorff measure that is used in obtaining (\ref{Snass1}) is its scaling property.

Let $\nu$ be some probability measure on $\R^{2D}$ which is absolutely continuous with respect to the $d$-dimensional Hausdorff measure ${\cal H}^d$ on $\R^{2D}$. Hence, it is supported on a fractal set of dimension $d$, and can be written
\beq
	d\nu(\underline{x}) = f(\underline{x})\,d{\cal H}^d(\underline{x})
\eeq
where $\underline{x}\in \R^{2D}$. Note that the Hausdorff measure satisfies
\beq
	d{\cal H}^d(s \underline{x} + \underline{y}) =
	s^d\,d{\cal H}^d(\underline{x})
\eeq
for all $\underline{x}, \underline{y} \in \R^{2D}$, $s>0$. Let us assume that $\lim_{s\to0} {\rm sup}(f(\underline{x}):|\underline{x}|<s)=M$ is finite. This implies that, if $F$ is a positive function on $\R^{2D}$ that decays fast enough at infinity, then
\beq\label{bound}
	0\leq\int d\nu(\underline{x})\,F(\underline{x}/s) = s^d
	\int d{\cal H}^d(\underline{x})
	\,f(s\underline{x})\,F(\underline{x}) \leq
	K s^d
\eeq
as $s\to0$, where $K = M \int d{\cal H}^d(\underline{x})\,F(\underline{x})$ is finite.

Let $\cvec v\in \CP^D$ (homogeneous coordinates), and let $U_{\cvec v}\subset \R^{2D}$ be a coordinate patch around $\cvec v$ (which can be see as an open subset of the tangent plane at $\cvec v$). Let $\phi_{\cvec v}:U_{\cvec v}\to\CP^D$ be the associated coordinate diffeomorphism, with $\phi_{\cvec v}(0) = \cvec v$. Given any $\cvec w\in \phi_{\cvec v}(U_{\cvec v})$, we can define $\underline{x}=\phi_{\cvec v}^{-1}(\cvec w)\in U_{\cvec v}$. We normalize $\phi_{\cvec v}$ in such a way that if $\cvec w$ and $\h{\cvec w}$ both tend to $\cvec v$, then
\beq\label{eqq1}
	D(\cvec w,\h{\cvec w}) \sim |\underline{x}-\h{\underline{x}}|.
\eeq

Let $\mu$ be a probability measure on $\CP^D$ such that for every $\cvec w\in \CP^D$ in a neighborhood of $\cvec v$, there is a measure $\nu_{\cvec v}$ as above such that
\beq\label{eqq2}
	d\mu(\cvec w) = d\nu_{\cvec v}(\phi_{\cvec v}^{-1}(\cvec w)) = d\nu_{\cvec v}(\underline{x}).
\eeq
This is a measure supported on a fractal set on $\CP^D$ (of fractal dimension $d$), and by integration over a fractal set on $\CP^D$, we mean integration over such a measure.

Using (\ref{Sn}), (\ref{replacement}) (again noting that this holds for every measure $\mu$ in (\ref{psiO})), (\ref{eqq1}) and (\ref{eqq2}), we have
\begin{eqnarray}
	S_n &\sim & \frc1{1-n} \log\lt[
	\int \lt(\prod_{\alpha=1}^n d\mu(\cvec v_\alpha)\rt)
	\lt( \prod_{\alpha=1}^n e^{-\frc{m}8 D(\cvec v_{\alpha},\cvec v_{\alpha+1})} \rt)
	\rt] \n
	&\sim& \frc1{1-n} \log\lt[\int d\mu(\cvec v_1)
	\int \lt(\prod_{\alpha=2}^n d\nu_{\cvec v_1}(\underline{x}^\alpha)\rt)
	\lt( \prod_{\alpha=2}^n e^{-\frc m8 \lt[\sum\limits_{\alpha=2}^{n-1}
	|\underline{x}^{\alpha+1}-\underline{x}^{\alpha}|^2+|\underline{x}^2|^2 + |\underline{x}^n|^2\rt]} \rt)
	\rt].\no
\end{eqnarray}
Since $\mu$ is a probability measure (hence has a finite integration), it must be that for almost all $\cvec v$ with respect to $\mu$, the measure $d\nu_{\cvec v}(\underline{u})$ has an associated density $f(\underline{u})$ that is finite in a neighbourhood of $\underline{u}=0$. Hence, we can use (\ref{bound}) (multiple integral generalization) to find
\[
	0\leq
	\int \lt(\prod_{\alpha=2}^n d\nu_{\cvec v_1}(\underline{x}^\alpha)\rt)
	\lt( \prod_{\alpha=2}^n e^{-\frc m8 \lt[\sum\limits_{\alpha=2}^{n-1}
	|\underline{x}^{\alpha+1}-\underline{x}^{\alpha}|^2+|\underline{x}^2|^2 + |\underline{x}^n|^2\rt]} \rt) \leq K(\cvec v_1)\,m^{\frac{d(1-n)}{2}}
\]
where $K(\cvec v_1)$ is almost surely finite. In order to arrive at the claimed asymptotic behaviour, we must guarantee that there is a subset of $\mu$-measure nonzero for $\cvec v_1$ such that the lower bound for the above integral is strictly greater than 0, and further of the order $m^{\frc{d(1-n)}2}$. In the context of (\ref{bound}), this occurs if the limit $\lim_{s\to 0}f(s\underline{x})=\t{f}(\underline{x})$ is integrable and integrates to a nonzero value. It seems natural to believe that this holds under $\nu_{\cvec v}$-integration for almost all $\cvec v$ with respect to $\mu$; it in fact holds for densities $f(\underline{x})$ that are continuous on their support. Assuming this lower bound property, we then find
\beq
	S_n\sim \frc{d}2 \log m
\eeq
in agreement with (\ref{Snass1}).
\section{Conclusions and outlook}
In this paper we have studied the von Neumann and R\'enyi bi-partite entanglement entropies in the thermodynamic limit for spin-$s$ quantum systems. This includes quantum spin chains as well as any other quantum many-body systems whose Hilbert space is described by a tensor product of spin-$s$ sites. In our study, we have considered a very special type of quantum states known as permutation symmetric states. These are states which are permutation symmetric under the exchange of any sites. In the present context, this type of states was considered in \cite{pop2} and part of our work has been devoted to extending the results of this work and to set them in a more general context.

The motivation to study the entropy of permutation symmetric states is two-fold: on the one hand this kind of states, for finite chains, are closely related to Dicke states, which are widely studied in quantum computation and have been experimentally realized \cite{exp}. On the other hand, the behaviour of the entropy for large subsets $A$, as defined in (\ref{vneu}), is unusual and not dictated by conformal invariance. As for conformal critical points, logarithmic scaling is found. However, unlike in the latter theories, the numerical coefficient of the logarithm is not related to a central charge and is the same for both the von Neumann and R\'enyi entropies (e.g. $n$-independent). Hence in particular, it is also the same for the so-called single-copy entropy (the limit $n\to\infty$ of the R\'enyi entropy). In this paper, following on previous work \cite{permutation,fractal} for the spin-$\frac{1}{2}$ case, we have shown that the coefficient of the leading logarithmic behaviour has a geometric meaning; it measures the geometric dimension of the set of basic zero-entropy states in terms of which the permutation symmetric states can be expressed. For spin-$\frac{1}{2}$ the geometries involved are subsets of the Bloch sphere. For spin-$s$ we have shown that this generalises to $\mathbb{C}\mathrm{P}^{2s}$.

From a mathematical standpoint the most challenging aspect of our work has been to provide a meaningful prescription which would allow us to carry out the
thermodynamic limit, that is when the length of the chain $N$ and the dimension of the Hilbert space tend to infinity. We have provided plausible arguments as well as examples tending to prove the existence of a probability measure $\mu$ supported on all subsets of $\mathbb{C}\mathrm{P}^{2s}$ with respect to the Study-Fubini metric (the natural metric in $\mathbb{C}\mathrm{P}^{2s}$). Provided this measure exists (Theorem 1), we have proven that correlation functions of the infinite chain as well as its von Neumann and R\'enyi entropies may always be expressed as integrals over such measure in a very general form (Theorem 2).
 
When considering the thermodynamic limit of permutation symmetric states we have in particular considered a very particular basis for such states, namely that previously introduced in the work of Popkov et al.~\cite{pop2} as a generalization of their previous work \cite{pop1} for the spin-$\frac{1}{2}$ case. In this basis, which we referred to as the elementary vectors or states, permutation symmetric states are characterized by fixed ratios $p_i$ of spins with $S_z$-projection given by $s-i$ and $i=0,1,\ldots,D$. In \cite{pop2} the logarithmic scaling behaviour of the von Neumann entropy for these states had already been found. Here we have reproduced and extended that result to the R\'enyi entropy, giving the scaling behaviours (\ref{renyi2}) and (\ref{vn3}). Popkov et al.~were able to obtain their results by exploiting the particularly nice combinatoric features of their chosen state such as the fact that non vanishing eigenvalues of the density matrix are generalised binomial coefficients. For large blocks, a saddle point analysis of these coefficients then leads to the behaviour (\ref{vn3}). In our case this behaviour follows from Theorem 2, which gives our particular formula for the entanglement entropy based on twist operators \cite{permutation}. Further, the elementary vectors are very particular linear combinations of our zero-entropy vectors, and thanks to the generality of this formula we have been able to investigate the entanglement entropy of much more general linear combinations (both finite and infinite) and we have shown that the scaling behaviour of the entropy will generally be different depending on the chosen state. Besides the case above, we have explicitly computed this behaviour for the state described in Subsection 5.3, giving (\ref{vn2}). In this case, the coefficient of the logarithmic term is $2s$ and is maximal, meaning that no linear combination of permutation symmetric states can have a faster growing entropy than this particular one. We have also explained how other states may be systematically constructed giving coefficients $\frac{d}{2}=0,\frac{1}{2},1,\frac{3}{2},\ldots,2s$ for the logarithmic divergency, and how in these cases the $O(1)$ correction term to this divergency has an interpretation in the context of geometric quantum mechanics.
Finally, we have shown how one may construct states for which the quantity $\frac{d}{2}$ takes irrational values, with $d$ representing the fractal dimension of some fractal subset of $\mathbb{C}\mathrm{P}^{2s}$. We have provided an explicit construction for $s=\frac{1}{2}$ and the Cantor set and a general argument for higher spins and general fractal sets.

We leave to future work a more extensive discussion and proof of Theorem 1, that is the existence of a measure $\mu$ with the given properties. A rigorous proof of this statement turns out to be quite involved and is closely related to the famous moment problem in mathematics which asks the question: given a set of moments $x_0, x_1,x_2,\ldots$ under which conditions is it possible to find a measure $\mu$ such that $x_n=\int x^n d\mu $? A general answer to this question in the higher-dimensional case is not known; however in our particular application we believe that such answer can be found. 

It would be interesting to investigate further the precise relationship between our results and the geometric entropy defined in the context of geometric quantum mechanics. Also, a numerical test of our findings would be of interest.

\vspace{1cm}
\noindent\textbf{{Acknowledgement}}: We would like to thank Denis Bernard for discussions. We are indebted to the Galileo Galilei Institute for Theoretical Physics for hospitality and to the INFN for partial support during the completion of this work.

\bibliographystyle{phreport}

\small

\begin{thebibliography}{10}

\bibitem{specialissue}
J.~Calabrese, P.~Cardy and D.~B. (ed),
\newblock {Entanglement entropy in extended quantum systems},
\newblock J. Phys. A {\bf 42}, 500301 (2009).

\bibitem{HolzheyLW94}
C.~Holzhey, F.~Larsen, and F.~Wilczek,
\newblock Geometric and renormalized entropy in conformal field theory,
\newblock Nucl. Phys. {\bf B424}, 443--467 (1994).

\bibitem{Calabrese:2004eu}
P.~Calabrese and J.~L. Cardy,
\newblock Entanglement entropy and quantum field theory,
\newblock J. Stat. Mech. {\bf 0406}, P002 (2004).

\bibitem{Calabrese:2005in}
P.~Calabrese and J.~L. Cardy,
\newblock Evolution of entanglement entropy in one-dimensional Systems,
\newblock J. Stat. Mech. {\bf 0504}, P010 (2005).

\bibitem{random}
G.~Refael and J.~E. Moore,
\newblock {Criticality and entanglement in random quantum systems},
\newblock J. Phys. A {\bf 42}, 504010 (2009).

\bibitem{pop1}
V.~Popkov and M.~Salerno,
\newblock Logarithmic divergence of the block entanglement entropy for the
  ferromagnetic Heisenberg model,
\newblock Phys. Rev. A {\bf 71}, 012301 (2005).

\bibitem{pop2}
V.~Popkov, M.~Salerno, and G.~Schutz,
\newblock Entangling power of permutation-invariant quantum states,
\newblock Phys. Rev. A {\bf 72}, 032327 (2005).

\bibitem{vidal}
J. I. Latorre, R. Or\'us, E. Rico and J. Vidal,
\newblock Entanglement entropy in the Lipkin-Meshkov-Glick model,
Phys. Rev. A {\bf 71}, 064101 (2005).

\bibitem{ravanini}
E.~Ercolessi, S.~Evangelisti, F.~Franchini, and F.~Ravanini,
\newblock {Essential singularity in the R\'enyi entanglement entropy of the
  one-dimensional XYZ spin-1/2 chain},
\newblock Phys.Rev. {\bf B83}, 012402 (2011).

\bibitem{permutation}
O.~A. Castro-Alvaredo and B.~Doyon,
\newblock {Permutation operators, entanglement entropy, and the XXZ spin chain
  in the limit $\Delta \rightarrow -1$},
\newblock J. Stat. Mech. {\bf 1102}, P02001 (2011).

\bibitem{fractal}
O.~A. Castro-Alvaredo and B.~Doyon,
\newblock {Entanglement entropy of highly degenerate states and fractal
  dimensions},
\newblock Phys. Rev. Lett. {\bf 108}, 120401 (2012).

\bibitem{SC1}
J. Eisert and M. Cramer, Phys. Rev. {\bf A72}, 042112 (2005).

\bibitem{SC2}
R. Or\'us, J. I. Latorre, J. Eisert and M. Cramer, Phys. Rev. {\bf A73}, 060303(R) (2006).


\bibitem{dicke}
R.~H. Dicke,
\newblock Coherence in Spontaneous Radiation Processes,
\newblock Phys. Rev. {\bf 93}, 99--110 (1954).


\bibitem{LMG}
H. J. Lipkin, N. Meshkov and A. J. Glick, Nucl. Phys. {\bf 62}, 188 (1965).

\bibitem{MGL}
N. Meshkov, A. J. Glick and H. J. Lipkin, Nucl. Phys. {\bf 62}, 199 (1965).

\bibitem{GLM}
A. J. Glick, H. J. Lipkin and N. Meshkov, Nucl. Phys. {\bf 62}, 211 (1965).

\bibitem{vidal2}
R. Or\'us, S Dusuel and J Vidal, Equivalence of critical scaling laws for many-body entanglement in the Lipkin-Meshkov-Glick model, Phys. Rev. Lett. {\bf 101}, 025701 (2008).

\bibitem{gqm1}
T.~W.~B. Kibble,
\newblock {Relativistic models of nonlinear quantum mechanics},
\newblock Commun. Math. Phys. {\bf 64}, 73--82 (1978).

\bibitem{gqm2}
T.~W.~B. Kibble,
\newblock {Geometrization of quantum mechanics},
\newblock Commun. Math. Phys. {\bf 65}, 189--201 (1979).

\bibitem{gqm3}
D.~C. Brody and L.~P. Hughston,
\newblock {Geometric quantum mechanics},
\newblock J. Geom. Phys. {\bf 38}, 19--53 (2001).

\bibitem{fubini}
G.~Fubini,
\newblock {Sulle metriche definite da una forme Hermitiana},
\newblock Atti. Istit. Veneto {\bf 63}, 502--513 (1904).

\bibitem{study}
E.~Study,
\newblock {K\"urzeste Wege im komplexen Gebiet},
\newblock Math. Ann. {\bf 60}, 321--378 (1905).

\bibitem{sfbook1}
S.~Kobayashi and K.~Nomizu,
\newblock {Foundations of Differential Geometry, Vol. 2, Wiley, New York},
\newblock (1969).

\bibitem{sfbook2}
V.~I. Arnold and K.~Nomizu,
\newblock {Mathematical Methods of Classical Mechanics, 2nd ed.,
  Springer-Verlag,},
\newblock (1989).

\bibitem{bures}
D.~Bures,
\newblock {An extension of Kakutani's theorem on infinite product measures to
  the tensor product of semifinite $w^*$-algebras},
\newblock Trans. Am. Math. Soc. {\bf 135}, 199--212 (1969).

\bibitem{uhlmann}
A.~Uhlmann,
\newblock {The Metric of Bures and the Geometric Phase, Quantum Groups and
  Related Topics: Proceedings of the First Max Born Symposium},
\newblock (1992).

\bibitem{hc}
W.~C. Graustein,
\newblock {Introduction to Higher Geometry, Macmillan, New York},
\newblock pages 29--49 (1930).

\bibitem{chaos}
H.~Peitgen, H.-O.~Juergens and D.~Saupe,
\newblock {Chaos and Fractals: New Frontiers of Science, Springer Verlag, New
  York},
\newblock (2004).

\bibitem{exp}
R.~Prevedel, G.~Cronenberg, M.~S. Tame, M.~Paternostro, P.~Walther, M.~S. Kim,
  and A.~Zeilinger,
\newblock Experimental Realization of Dicke States of up to Six Qubits for
  Multiparty Quantum Networking,
\newblock Phys. Rev. Lett. {\bf 103}, 020503 (2009).

\end{thebibliography}

\end{document}